\documentclass[twocolumn]{article}
\usepackage[a4paper, margin=1.8cm]{geometry}

\usepackage{amsmath,amssymb,amsfonts}
\usepackage{amsthm}
\usepackage[hidelinks]{hyperref}
\usepackage{orcidlink}
\usepackage{graphicx}
\usepackage{xcolor}
\usepackage{enumitem}
\usepackage{caption}
\usepackage{algorithm, algorithmic}
\usepackage{lipsum}
\usepackage{float}
\usepackage{array}
\usepackage{booktabs}
\usepackage{multirow}
\usepackage{siunitx}
\usepackage{url}
\usepackage{mathrsfs}
\usepackage{ragged2e}
\usepackage{tikz}
\usepackage{pgfplots}
\usepackage{comment}
\usepackage{bm}

\usetikzlibrary{plotmarks, fit}

\pgfplotsset{
    compat=1.18,
    every axis/.append style={font=\small, tick scale binop=\times, /pgf/number format/1000 sep={}},
    minor grid style={dash pattern=on 1.5pt off 1pt},
    legend image code/.append code={
        \node[fit=(current bounding box),inner xsep=0.1em, inner ysep=0.1em]{};
    },
    every crossref picture/.style={baseline,yshift=0.3em},
}

\newcommand{\tikzref}[1][]{%
    \begin{tikzpicture}[baseline,yshift=0.3em]
        \draw [mark repeat=2,mark phase=2,#1]
        plot coordinates {
            (0cm,0cm)
            (0.25cm,0cm)
            (0.5cm,0cm)
        }; \node[fit=(current bounding box),inner xsep=0.1em]{};
    \end{tikzpicture}%
}

\newcommand{\tikzreffill}[1][]{%
    \begin{tikzpicture}[baseline,yshift=0em]
        \fill[#1] (0,0) rectangle (1em,0.6em);
        \node[fit=(current bounding box),inner xsep=0.1em]{};
    \end{tikzpicture}%
}

\DeclareCaptionLabelSeparator{periodquad}{.\quad}

\newcolumntype{L}[1]{>{\raggedright\let\newline\\\arraybackslash\hspace{0pt}\linespread{1}\selectfont}m{#1}}
\newcolumntype{C}[1]{>{\centering\let\newline\\\arraybackslash\hspace{0pt}\linespread{1}\selectfont}m{#1}}
\newcolumntype{R}[1]{>{\raggedleft\let\newline\\\arraybackslash\hspace{0pt}\linespread{1}\selectfont}m{#1}}
\newcolumntype{J}[1]{>{\linespread{1}\selectfont}m{#1}}

\newcommand{\TopRule}{\specialrule{1pt}{0pt}{2pt}}
\newcommand{\BottomRule}{\specialrule{1pt}{2pt}{0pt}}
\newcommand{\MidRule}{\specialrule{0.5pt}{1pt}{1pt}}

\DeclareMathOperator*{\argmin}{arg\,min}
\DeclareMathOperator*{\diag}{diag}

\DeclareMathOperator*{\inter}{int}
\DeclareMathOperator*{\conv}{conv}
\newcommand{\norm}[1]{\left\lVert #1 \right\rVert}

\newtheorem{theorem}{Theorem}

\newtheorem{proposition}{Proposition}

\newtheorem{remark}{Remark}
\newtheorem{assumption}{Assumption}

\allowdisplaybreaks

\definecolor{spice-blue}{rgb}{0, 0, 1}
\definecolor{spice-cyan}{rgb}{0, 0.75, 0.75}
\definecolor{spice-green}{rgb}{0, 0.5, 0}
\definecolor{spice-red}{rgb}{1, 0, 0}
\definecolor{spice-brown}{rgb}{0.5, 0.25, 0}
\definecolor{spice-orange}{rgb}{1, 0.5, 0}

\newcommand{\quotes}[1]{``#1''}

\usepackage{etoolbox}
\AtBeginEnvironment{minipage}{\captionsetup{justification=centering}}

\begin{document}

\title{A Fully Analog Implementation of Model Predictive Control With Application to Buck Converters}
\date{}

\author{
    Simone Pirrera\thanks{
        S. Pirrera and D. Regruto are with the Department of Control and Computer Engineering, Politecnico di Torino, 10129 Turin, Italy (e-mail: simone.pirrera@polito.it; diego.regruto@polito.it).},
    Lorenzo Calogero\thanks{
        L. Calogero was with the Department of Electronics and Telecommunications, Politecnico di Torino, 10129 Turin, Italy. He is now with Argotec S.r.l., 10099 San Mauro Torinese, Italy (e-mail: lorenzo.calogero@polito.it).},
    Francesco Gabriele\thanks{
        F. Gabriele was with the Department of Electronics and Telecommunications, Politecnico di Torino, 10129 Turin, Italy. He is now with CEMSE, King Abdullah University of Science and Technology, Thuwal 23955, Saudi Arabia (e-mail: francesco.gabriele@kaust.edu.sa).}, \\
    Diego Regruto$^*$, Alessandro Rizzo\thanks{
        A. Rizzo is with the Department of Electronics and Telecommunications, Politecnico di Torino, 10129 Turin, Italy (e-mail: alessandro.rizzo@polito.it).}, and Gianluca Setti\thanks{
        G. Setti was with CEMSE, King Abdullah University of Science and Technology, Thuwal 23955, Saudi Arabia. He is now with the Department of Electronics and Telecommunications, Politecnico di Torino, 10129 Turin, Italy (e-mail: gianluca.setti@polito.it).}
}

\maketitle

\begin{abstract}
    This paper proposes a novel approach to design analog electronic circuits that implement Model Predictive Control (MPC) policies for dynamical systems described by affine models. Effective approaches to define a reduced-complexity explicit MPC are combined and applied to realize an analog circuit comprising a limited set of low-latency, commercially available components. The practical feasibility and effectiveness of the proposed approach are demonstrated through its application in the design of a novel controller for DC-DC Buck converters. We formally analyze the stability of the resulting system and conduct extensive numerical simulations to demonstrate the control system's performance in rejecting line and load disturbances.
\end{abstract}


\section{Introduction}

Model predictive control (MPC) is one of the most widely adopted control techniques today, due to its flexibility to adapt to plants of diverse nature and, at the same time, to handling constraints \cite{faulwasser2021recent}.
Despite these desirable features, the practical implementation of MPC is not straightforward. First, the typical implementation of MPC imposes a significant overhead in terms of costs and circuit area due to the need to employ expensive digital hardware, including analog-to-digital converters (ADC), digital-to-analog converters (DAC), and a computing device, making the overall system too expensive for mass production, where a limited-budget design is often a primary requirement. Second, despite rapid advances in the development of novel algorithms for convex optimization~\cite{paper-calogero-2024,tac_cmo_25}, the need to solve quadratic programming (QP) problems at each iteration renders the solution infeasible for systems exhibiting fast dynamics. 

A widely adopted methodology, known for achieving significant computational efficiencies, involves the formulation of an explicit MPC (EMPC) control law. As demonstrated in~\cite{bemporad2000explicit,paper-bemporad-2002}, a piecewise affine (PWA) function can be computed offline and used to define the MPC control input, obviating the need for online optimization. Although several algorithms have been proposed to efficiently compute the EMPC control law~\cite{patrinos2010new,OBERDIECK2017103,bemporad2015multiparametric}, the EMPC approach still raises criticalities in budget-constrained and fast-sampled applications. On the one hand, the need for digital hardware persists; on the other hand, when the PWA function is defined by several regions, its real-time evaluation may remain critical: the parameters defining the function may exceed the hardware storage capacity, and identifying the region containing the current parameter may be computationally expensive~\cite{paper-bemporad-2021}. These issues are reported, e.g., in~\cite{wisniewski_explicit_2021}, which applies EMPC to Buck converters.
A viable alternative is brought by custom hardware implementations. Notable results in this direction are~\cite{WillsTCST12_fpga} and~\cite{vichik_solving_2014}. In \cite{WillsTCST12_fpga} the authors use a field-programmable gate array to implement an active-set algorithm to solve the QP, but its performance remains limited by the need to perform analog-to-digital and digital-to-analog conversions. On the other hand, \cite{vichik_solving_2014} constructs an analog circuit that solves QP, thereby making it suitable for MPC implementations. However, the applicability of this method is limited by the resulting circuit complexity. 

Here, we present a novel, general methodology for designing a fully analog electronic circuit that implements MPC controllers for affine systems. First, we derive an explicit MPC control law and apply state-of-the-art approaches to describe the controller in terms of a limited number of regions. Next, we define the circuit implementation using commercially available low-latency components: comparators, resistors, operational amplifiers (OP-AMPs), and one multiplexer (MUX). 

Furthermore, the effectiveness and feasibility of our approach are validated through the application to the control of DC-DC Buck converters. Buck converters are devices that regulate a supply voltage to a lower, constant value. Buck converters are used in various applications, including the biomedical and renewable energy fields; see, e.g.,~\cite{chen_analysis_2013,pavlovsky2013assessment,park2023off}.
Accurately regulating the output voltage in Buck converters is fundamental for the proper operation of the devices supplied by the converter. However, the tracking accuracy of the converter is hindered by unpredictable and sudden variations of the input supply voltage (line disturbances) and of the current drawn by the load (load disturbances)~\cite{book-erickson-2020,gab25_estimator}. 
To improve tracking performance, several approaches have been considered. As to line disturbance attenuation, we mention the popular feedforwarding technique~\cite{Amer_Access_2024}. In contrast, the problem of attenuating the effects of load disturbances is more challenging. It has been addressed by methods based on the design of circuitry to modify either the feedback signal or the pulse width of the Pulse Width Modulated (PWM) signal in response to load variations; see~\cite{He_ISCAS_2022, Hsu_TCASII_2019} for recent works in this direction. However, these solutions lack solid theoretical foundations. Other approaches aim at estimating the output current disturbance to apply a proper feedforward action. Among these, we mention the extended state observer~\cite{korompili_linear_2025}, generalized proportional-integral observer~\cite{wang_gpi_2019}, and the load estimator-compensator (LEC)~\cite{gab25_estimator} approaches. These solutions are often combined with optimal control design:~\cite{korompili_linear_2025} uses linear quadratic gaussian control in conjunction with the extended state observer, and~\cite{gab25_estimator} uses $\mathcal{H}_\infty$-optimal voltage mode control (VMC) in conjunction with the LEC.

For Buck converters, the MPC controller design approach is particularly compelling due to its ability to directly handle the limitation imposed by input duty-cycle saturation between \num{0}\% and \num{100}\%. 
The application of MPC to Buck converters has been explored in several contributions, e.g.,~\cite{cim25,mariethoz_comparison_2010,liu_fast_2018,albira_adaptive_2021}. These works report successful results in practical implementations, but, in all cases, the sampling frequency must be kept low to ensure convergence of the optimization algorithm. More specifically,~\cite{cim25,mariethoz_comparison_2010}, and~\cite{albira_adaptive_2021} report switching frequencies limited to~\SI{10}{\kilo\hertz},~\SI{20}{\kilo\hertz}, and~\SI{25}{\kilo\hertz}, respectively. Similar considerations apply to~\cite{liu_fast_2018}, which implements the MPC using a field-programmable gate array. 

To apply the proposed approach to DC-DC Buck converters, we first introduce a discrete-time (DT) mathematical model of the device and its linearization. Next, to ensure proper load-disturbance rejection performance, we use the load estimator proposed in~\cite{gab25_estimator}, which enables a low-cost analog implementation. Finally, we apply the proposed analog EMPC design: thanks to the adopted region-reduction strategies, we observe a significant improvement over previous works on EMPC design for the Buck (see~\cite{wisniewski_explicit_2021}), thus ensuring a low-cost circuit design and a faster implementation compared to the approach proposed in~\cite{vichik_solving_2014}. Furthermore, we provide a formal proof of the stability of the feedback control system's equilibrium point, robustly with respect to disturbances.

Our controller is validated through a comprehensive simulation study to demonstrate the feasibility and effectiveness of the proposed approach. We analyze system performance under parametric uncertainty in Buck component values using Monte Carlo simulations, considering both ceramic and electrolytic capacitor cases. Next, we perform an accurate low-level simulation utilizing the manufacturer's models of commercially available components, demonstrating the feasibility of the proposed method even in the presence of component non-idealities. The proposed analog implementation enables high-frequency sampling at \SI{500}{\kilo\hertz}, which is compatible with modern Buck designs and considerably improves upon previous solutions reported in~\cite{cim25,mariethoz_comparison_2010,liu_fast_2018,albira_adaptive_2021}. Our results indicate that the system exhibits outstanding disturbance-rejection performance, outperforming standard methods.

\subsection{Outline}
Sec.~\ref{sec:probl_formul} reviews the standard formulation of the MPC problem. Sec.~\ref{sec:mpc_design} expands on the proposed design method by proposing the EMPC formulation and its analog circuit implementation. In Sec.~\ref {sec:buck_applic} we apply the proposed approach to the DC-DC Buck converter control problem: we model the system, detail how to estimate unmeasurable quantities, and provide stability guarantees. Sec.~\ref{sec:results} depicts the obtained numerical results. Finally, Sec.~\ref{sec:conclusion} draws conclusions.

\subsection{Notation}
In the following, $x = [x_i]_{i=1}^N \in \mathbb{R}^{nN}$ is the vertical concatenation of the vectors $x_i \in \mathbb{R}^n$; $I_n \in \mathbb{R}^{n \times n}$ is the identity matrix; $\bm{0}_n, \bm{1}_n \in \mathbb{R}^{n}$ are the null and all-$1$ vectors, respectively; $\|x\|_M \doteq \|M^{1/2}x\|_2 = \sqrt{x^\top Mx}$ is the weighted norm of vector $x$, with weighting matrix $M$, and $\|A\|_M = \|M^{1/2} A M^{-1/2}\|_2$ is the induced weighted norm of matrix $A$; $\otimes$ is the Kronecker product; $e_{n}^{(m)} \in \mathbb{R}^{n}$ is the $m$-th vector of the standard Euclidean basis of $\mathbb{R}^{n}$; $E_{n}^{(m)} \in \mathbb{R}^{n \times n}$ is the matrix with $1$s on the $m$-th subdiagonal and $0$s elsewhere; $\odot$, $\oplus$, and $\overline{\,\star\,}$ are the Boolean (logic) AND, OR, and NOT operators, respectively.

\section{MPC Problem Formulation} \label{sec:probl_formul}
This section reviews the MPC formulation that we will consider throughout the remainder of the paper. 
Let us consider an $n$-dimensional discrete-time (DT) affine system, i.e.,
\begin{subequations} \label{eq:sys_dt_lti}
    \begin{align}
        x_{k+1} &= A x_{k} + B u_k + B_{\nu} \nu_k + b, \\
        y_k &= C x_{k} + D u_k + D_\nu \nu_k + d,
    \end{align}
\end{subequations}
where $x \in \mathbb{R}^{n}$, $u \in \mathbb{R}^{n_u}$, and $y \in \mathbb{R}^{n_y}$ are the state, input, and output vectors, respectively, $\nu \in \mathbb{R}^{n_\nu}$ is a vector of exogenous disturbances, $A \in \mathbb{R}^{n \times n}$, $B \in \mathbb{R}^{n \times n_u}$, $B_\nu \in \mathbb{R}^{n \times n_\nu}$, $C \in \mathbb{R}^{n_y \times n}$, $D \in \mathbb{R}^{n_y \times n_u}$, and $D_\nu \in \mathbb{R}^{n_y \times n_u}$ are matrices describing the system dynamics, $b \in \mathbb{R}^{n}$ and $d \in \mathbb{R}^{n_y}$ are constant affine terms. 

We define, for each $k \geq 0$, the MPC optimal control problem for system~\eqref{eq:sys_dt_lti} as:
\begin{subequations} \label{eq:mpc_op}
    \begin{align}
        \min_{\hat{x},\hat{u},\hat{y}} & \sum_{i=0}^{N_p-1} \! \Big(\| \hat{y}_{i} - y_r \|_{Q}^2 + \| \hat{u}_{i} - u_r \|_{R}^2\Big) + \!\! \sum_{i=1}^{N_p-1} \! \| \hat{u}_{i} - \hat{u}_{i-1} \|_{R_\Delta}^2, \label{eq:mpc_op_a} \\
        \textrm{s.t.} \;\;\; & \hat{x}_{0} = x_k, \quad \hat{x}_{i+1} = A \hat{x}_{i} + B \hat{u}_{i} + B_{\nu}\nu_k + b, \label{eq:mpc_op_b} \\
        & \hat{y}_{i} = C \hat{x}_{i} + D \hat{u}_{i} + D_{\nu}\nu_k + d, \label{eq:mpc_op_c} \\
        & H_x \hat{x}_i \leq h_x, \quad H_u \hat{u}_i \leq h_u,~~i = 0, \ldots, N_p-1. \label{eq:mpc_op_d} 
    \end{align}
\end{subequations}
Here, the decision variables are the predicted state trajectory $\hat{x} = [\hat{x}_{i}]_{i=1}^{N_p} \in \mathbb{R}^{n N_p}$, the predicted output trajectory $\hat{y} = [\hat{y}_{i}]_{i=0}^{N_p-1} \in \mathbb{R}^{n_y N_p}$, and the input sequence $\hat{u} = [\hat{u}_i]_{i=0}^{N_p-1} \in \mathbb{R}^{n_u N_p}$, where $N_p$ is the prediction horizon.

The cost function~\eqref{eq:mpc_op_a} is composed of three terms: $ \| \hat{y}_{i} - y_r \|_{Q}^2 $ and $ \| \hat{u}_{i} - u_r \|_{R}^2 $ serve as regulation terms for the predicted output $\hat{y}_{i}$ and input $\hat{u}_{i}$ towards the constant reference output $y_r \in \mathbb{R}^{n_y}$ and input $u_r \in \mathbb{R}^{n_u}$, respectively, while $\| \hat{u}_{i} - \hat{u}_{i-1} \|_{R_{\Delta}}^2$ penalizes the variation in time of the inputs to obtain smoother predicted trajectories (see, e.g.,~\cite{paper-calogero-2023, paper-calogero-2025}). $Q, R_{\Delta} \succeq 0$ and $R \succ 0$ are symmetric weighting matrices of suitable dimensions.
We consider the following standard assumption on the reference $(u_r, y_r)$:
\begin{assumption} \label{ass:mpc_ref_equil}
    The triple $(u_r, x_r, y_r)$ is an equilibrium of the undisturbed system~\eqref{eq:sys_dt_lti}, i.e., with $\nu_k = \bm{0}_{n_\nu}$ it holds that
    \begin{subequations}
        \begin{align}
            x_r &= Ax_r + Bu_r + b, \\
            y_r &= C x_r + Du_r + d.
        \end{align}
    \end{subequations}
\end{assumption}

The equality constraints~\eqref{eq:mpc_op_b},~\eqref{eq:mpc_op_c} serve as the MPC prediction model, which corresponds to model~\eqref{eq:sys_dt_lti}. 
Eq.~\eqref{eq:mpc_op_d} imposes linear inequality constraints: $N_{c_x} \in \mathbb{N}$ on the states defined by $H_x \in \mathbb{R}^{N_{c_x},n}$ and $h_x \in \mathbb{R}^{N_{c_x}}$, and $N_{c_u} \in \mathbb{N}$ on the input defined by $H_u \in \mathbb{R}^{N_{c_u},n_u}$ and $h_u \in \mathbb{R}^{N_{c_u}}$. 
The considered MPC formulation is general and can incorporate additional features such as robustness and safety constraints. Importantly, these do not affect the proposed analog implementation.

By assuming that all states and disturbances are available for measurement, we define the control law by using the one-step receding horizon policy, i.e., at each time instant $k$, the first optimal input $u_k^* \doteq \hat{u}_0^*$ given by the MPC problem~\eqref{eq:mpc_op} is applied to the plant~\eqref{eq:sys_dt_lti} over the time interval $I_k \doteq [kT,(k+1)T)$. This strategy implicitly defines a static state-feedback control policy $\pi: \mathbb{R}^{n_p} \to \mathbb{R}^{n_u}$, i.e.,
\begin{align} \label{eq:mpc_policy}
    u_k^* = \pi(p_k),
\end{align}
where $p_k \doteq [x_k^\top, \; \nu_k^\top]^\top \in \mathbb{R}^{n_p}$ is the vector of variables acting as parameters in problem~\eqref{eq:mpc_op}, with $n_p = n+n_\nu$. 

The MPC problem~\eqref{eq:mpc_op} can be rewritten in a more compact QP form, comprising only $\hat{u}$ as decision variables, i.e.,
\begin{subequations} \label{eq:mpc_qp}
    \begin{align}
        \min_{\hat{u}} \quad & \frac{1}{2}\hat{u}^\top H \hat{u} + (F p_k + c)^\top \hat{u} \label{eq:mpc_qp_a} \\
        \textrm{s.t.} \quad & G \hat{u} \leq w + Kp_k, \label{eq:mpc_qp_b}
    \end{align}
\end{subequations}
where $H \in \mathbb{R}^{n_u N_p \times n_u N_p}$, $F \in \mathbb{R}^{n_u N_p \times n_p}$, $c \in \mathbb{R}^{n_u N_p}$, $G \in \mathbb{R}^{q \times N_p n_u}$ and $w \in \mathbb{R}^q$. This transformation involves eliminating the constraints~\eqref{eq:mpc_op_b},~\eqref{eq:mpc_op_c}, and rewriting Eqs.~\eqref{eq:mpc_op_a} and~\eqref{eq:mpc_op_d} as functions of $\hat{u}$ and $p_k$ only; see, e.g.,~\cite{bemporad2000explicit}. The corresponding development is reported in Appendix~\ref{app:qp_formul}.

As shown in~\cite{paper-bemporad-2021}, the positive definiteness of $Q$, $R$, and $R_\Delta$ implies $H = H^\top \succ 0$. Hence, for any $p_k$, the MPC problem~\eqref{eq:mpc_qp} is strongly convex, thereby admitting a unique global optimum $\hat{u}^*$ (see, e.g.,~\cite{bertsekas_ed3}). Consequently, the policy $\pi$ is well-defined and uniquely maps $p_k$ to the control input $u_k^* = \hat{u}_0^*$, according to the receding horizon scheme.

\begin{proposition} \label{prop:mpc_unique_sol}
    Given Assumption~\ref{ass:mpc_ref_equil}, assume that $x_r$ and $u_r$ satisfy the constraints~\eqref{eq:mpc_op_d}, and let $p_k = [x_r^\top, \bm{0}_{n_\nu}^\top]^\top$. Then, problem~\eqref{eq:mpc_qp} is uniquely solved by $\hat{u}^* = \bm{1}_{N_p} \otimes u_r$.
\end{proposition}
\begin{proof}
    By the equivalence of problems~\eqref{eq:mpc_op} and~\eqref{eq:mpc_qp}, they share the same unique solution $\hat{u}^*$. Then, $\hat{x}_{i+1} = x_r$, $\hat{u}_i = u_r$, $i = 0, \ldots, N_p-1$, is the global minimizer of problem~\eqref{eq:mpc_op} since, by assumption, it is a feasible trajectory under the constraints~\eqref{eq:mpc_op_b} and achieves zero cost. Therefore, $\hat{u}^* = \bm{1}_{N_p} \otimes u_r$ uniquely solves problem~\eqref{eq:mpc_qp} with $p_k = [x_r^\top, \bm{0}_{n_\nu}^\top]^\top$.
\end{proof}

\begin{proposition}
    The reference output $y_r$ is an equilibrium of the closed-loop undisturbed system~\eqref{eq:sys_dt_lti},~\eqref{eq:mpc_policy}, i.e.,
    \begin{subequations}
        \begin{align}
            x_r &= Ax_r + B\pi([x_r^\top, \bm{0}_{n_\nu}^\top]^\top) + b, \\
            y_r &= C x_r + D\pi([x_r^\top, \bm{0}_{n_\nu}^\top]^\top) + d.
        \end{align}
    \end{subequations}
\end{proposition}
\begin{proof}
    This result is a direct consequence of Assumption~\ref{ass:mpc_ref_equil} and Proposition~\ref{prop:mpc_unique_sol}.
\end{proof}

\section{Proposed Design Approach} 
\label{sec:mpc_design}

This section proposes a methodology for implementing MPC as a fully analog circuit. We first exploit the EMPC closed-form formulation and apply complexity-reduction, then describe the corresponding circuit implementation.

\subsection{Reduced-Complexity EMPC Design}
\label{sec:explicit_mpc}

We start our design by explicitly representing the MPC policy~\eqref{eq:mpc_policy} in closed form as stated in the following theorem.

\begin{theorem}[Explicit MPC~\cite{paper-bemporad-2002}] \label{th:empc}
    Assume that the parameters $p_k$ lie in a convex polytope $\mathcal{P} \subseteq \mathbb{R}^{n_p}$. Then, the optimal solution $u_k^* = \hat{u}_0^*$ of the MPC problem~\eqref{eq:mpc_qp}, at each $k \geq 0$, is given by the policy $\pi$ in Eq.~\eqref{eq:mpc_policy}, where:
    \begin{enumerate}[label={(\roman*)}, leftmargin=*]
        \item $\pi: \mathcal{P} \to \mathcal{U}$ is a continuous piecewise affine (PWA) function, defined over $R$ regions $\mathcal{R}_i$, $i=1,\ldots,R$, i.e.,
        \begin{align} \label{eq:empc_pwa_fun}
            u_k^* = \pi(p_k) = K_i p_k + l_i \quad \textrm{if} \;\;\; p_k \in \mathcal{R}_i.
        \end{align}
        
        \item The regions $(\mathcal{R}_i)_{i=1}^R$ are full-dimensional polytopes with non-overlapping interiors, forming a partition of $\mathcal{P}$, i.e.,
        \begin{align} \label{eq:empc_regions_partition}
            \bigcup\nolimits_{i=1}^{R} \mathcal{R}_i = \mathcal{P}, \quad \inter(\mathcal{R}_i) \cap \inter(\mathcal{R}_j) = \emptyset, \;\; i \neq j.
        \end{align}
    \end{enumerate}
\end{theorem}

Replacing the MPC problem~\eqref{eq:mpc_qp} with the EMPC policy~\eqref{eq:mpc_policy} offers the significant advantage of being able to compute the optimal control input using a static function. 
As detailed in Sec.~\ref{sec:circuit}, the structure of Eq.~\eqref{eq:empc_pwa_fun} enables implementing the policy \eqref{eq:mpc_policy} using only static analog components, yielding quasi-instantaneous function evaluation.

Yet, the size and cost of the resulting analog circuit is directly influenced by the number of regions $R$ describing Eq.~\eqref{eq:empc_pwa_fun}. To minimize costs, it is essential to reduce them. 
As shown in~\cite{paper-grieder-2003}, $R$ equals the number of possible combinations of active constraints for problem~\eqref{eq:mpc_qp}. Thus, a bound on $R$ is
\begin{align}
    R \leq \sum_{i=0}^{n_u N_p} \binom{q}{i},
\end{align}
because at most $n_u N_p$ constraints can be active at the optimum. In most cases, $R$ is much smaller, as most of the constraint combinations are never active at optimality.
%
In the following, we illustrate how to reduce the
number of regions using four complexity
reduction strategies.

\subsubsection{Move Blocking Strategy}

The move blocking strategy~\cite{paper-cagienard-2007} reduces the complexity of the MPC problem~\eqref{eq:mpc_qp} by defining a linear map between the input sequence $\hat{u}$ and a smaller set of decision variables $u_c \in \mathbb{R}^{n_u N_c}$, $N_c < N_p$, as
\begin{align} \label{eq:mpc_move_block_map}
    \hat{u} = Tu_c, \quad T \in \mathbb{R}^{N_p n_u \times N_c n_u},
\end{align}
where the matrix $T$ is a lower-trapezoidal design parameter. A popular choice is
\begin{align} \label{eq:mpc_move_block_matrix}
    T = \begin{bmatrix}
        I_{N_c-1} & \bm{0} \\
        \bm{0} & \bm{1}_{N_p - N_c + 1}
    \end{bmatrix} \otimes I_{n_u},
\end{align}
so that the first $N_c-1$ samples of $\hat{u}$ are independent, while the samples from $N_c-1$ to $N_p-1$ are bound to the same value.
Eq.~\eqref{eq:mpc_move_block_map} can be directly replaced into the MPC problem~\eqref{eq:mpc_qp}, obtaining its reduced version with variables $u_c$, i.e.,
\begin{subequations} \label{eq:mpc_qp_moveblock}
    \begin{align}
        \min_{u_c} \quad & \frac{1}{2}u_c^\top (T^\top H T) u_c + (T^\top F p_k + T^\top c)^\top u_c \label{eq:mpc_qp_a_moveblock} \\
        \textrm{s.t.} \quad & G T u_c \leq w + K p_k. \label{eq:mpc_qp_b_moveblock}
    \end{align}
\end{subequations}
Notice that all the fundamental properties of the MPC formulation~\eqref{eq:mpc_qp}, such as its strong convexity, are preserved. Instead, the optimal solution is not. Still, $T$ and $N_c$ can always be tuned to ensure a negligible closed-loop performance decrease.  

The move blocking strategy directly reduces the number of regions $R$ in the EMPC policy~\eqref{eq:mpc_policy}. Indeed, the number of decision variables reduces from $N_p n_u$ in Eq.~\eqref{eq:mpc_qp} to $N_c n_u$ in Eq.~\eqref{eq:mpc_qp_moveblock}, yielding the tighter upper bound
\begin{align}\label{eq:reg_bnd_binomial}
    R \leq \sum\nolimits_{i=0}^{n_u N_c} \binom{q}{i} < \sum\nolimits_{i=0}^{n_u N_p} \binom{q}{i}.
\end{align}
For more details on this aspect, we refer the reader 
to~\cite{paper-alessio-2009}.

\subsubsection{Optimal Merging of Regions} \label{sec:empc_regions_merge}

In the EMPC policy~\eqref{eq:mpc_policy}, regions sharing the same affine control law can be merged together, provided that such a merging produces a convex polytope~\cite{paper-bemporad-2021}.
To this end,~\cite{paper-geyer-2008} proposes an optimal merging algorithm named non-disjoint optimal merging, that yields the minimal number of merged regions while allowing for possible overlaps. This latter aspect poses no issue when evaluating the simplified EMPC policy, as overlaps always share the same affine function. Importantly, allowing non-disjoint polytopes leads to solutions with fewer regions and fewer facets (i.e., fewer inequalities defining the region) compared to the case where we restrict to disjoint polytopes~\cite{paper-geyer-2008}. The non-disjoint optimal merging algorithm is implemented in the Multi-Parametric Toolbox (MPT) for \textsc{Matlab}~\cite{mpt3}.

\subsubsection{Hyperplane Separation of Saturated Regions} \label{sec:empc_sat_regions_sep}

The MPC formulation~\eqref{eq:mpc_op} typically accounts for lower and upper bounds on the input, i.e., Eq.~\eqref{eq:mpc_op_d} includes constraints of the kind $u_\mathrm{lb} \leq \hat u_i \leq u_\mathrm{ub}$ for some $u_\mathrm{lb},u_\mathrm{ub}\in \mathbb{R}^{n_u}$. In this case, the EMPC policy~\eqref{eq:mpc_policy} naturally leads to several regions where the control law is constantly equal to either $u_\mathrm{lb}$ or $u_\mathrm{ub}$. Henceforth, such regions are referred to as saturated regions; the other regions, in contrast, are called unsaturated regions.

Denoting with $I_\mathrm{lb}$ and $I_\mathrm{ub}$ the sets of indices of the regions saturated at the lower and upper bound, respectively, we define
\begin{align}
    \underline{\mathcal{S}} \doteq \bigcup\nolimits_{i\in I_\mathrm{lb}} \mathcal{R}_i, \quad \overline{\mathcal{S}} \doteq \bigcup\nolimits_{i\in I_\mathrm{ub}} \mathcal{R}_i.
\end{align}
It is worth noticing that the sets $\underline{\mathcal{S}}$ and $\overline{\mathcal{S}}$ are, in general, non-convex and, possibly, non-connected. Also, we define $I_\mathrm{unsat} \doteq \{1,\ldots,R\} \smallsetminus (I_\mathrm{lb} \cup I_\mathrm{ub})$ as the set of indices of the unsaturated regions, and the union of unsaturated regions as
\begin{align}
    \mathcal{R}_\mathrm{unsat} \doteq \bigcup\nolimits_{i \in I_\mathrm{unsat}} \mathcal{R}_i.
\end{align}

By Theorem~\ref{th:empc}, the continuity of the EMPC policy ensures that $\underline{\mathcal{S}} \cap \overline{\mathcal{S}} = \emptyset$. Also, since by Theorem~\ref{th:empc} the polytopes $\mathcal{R}_i$ do not overlap, it holds that $\mathcal{P} = \underline{\mathcal{S}} \cup (\mathcal{R}_\mathrm{unsat}) \cup \overline{\mathcal{S}}$, $\inter(\mathcal{R}_\mathrm{unsat}) \cap \inter(\underline{\mathcal{S}}) = \emptyset$, $\inter(\mathcal{R}_\mathrm{unsat}) \cap \inter(\overline{\mathcal{S}}) = \emptyset$. With this notation, the MPC policy in Eq.~\eqref{eq:mpc_policy} is rewritten as:
\begin{align} \label{eq:empc_policy_sat}
    \pi(p_k) = \begin{cases}
        K_i p_k + l_i & \textrm{if} \;\; p_k \in \mathcal{R}_i, \; i \in I_\mathrm{unsat}, \\
        u_\mathrm{lb} & \textrm{if} \;\; p_k \in \underline{\mathcal{S}}, \\
        u_\mathrm{ub} & \textrm{if} \;\; p_k \in \overline{\mathcal{S}}. \\
    \end{cases}
\end{align}
We reduce the complexity of Eq.~\eqref{eq:empc_policy_sat} by removing the saturated regions. This is achieved by introducing a function $\sigma: \mathcal{P} \to \mathbb{R}$, called \textit{separation function}, that separates the sets $\underline{\mathcal{S}}$ and $\overline{\mathcal{S}}$, according to its sign, i.e.,
\begin{align}
    \sigma(p) < 0, \;\; \forall\, p \in \underline{\mathcal{S}}, \quad \sigma(p) > 0, \;\; \forall \, p \in \overline{\mathcal{S}}.
\end{align}
Using $\sigma$, we can equivalently rewrite Eq.~\eqref{eq:empc_policy_sat} as
\begin{align} \label{eq:empc_policy_sat_sep}
    {\pi}(p_k) \! = \! \begin{cases}
        K_i p_k + l_i & \textrm{if} \;\; p_k \in \mathcal{R}_i, \; i \in I_\mathrm{unsat}, \\
        u_\mathrm{lb} & \textrm{if} \;\; p_k \notin \mathcal{R}_\mathrm{unsat}, \;\; \sigma(p_k) < 0, \\
        u_\mathrm{ub} & \textrm{if} \;\; p_k \notin \mathcal{R}_\mathrm{unsat}, \;\; \sigma(p_k) > 0. \\
    \end{cases}
\end{align}
The formulation~\eqref{eq:empc_policy_sat_sep} involves a reduced number of regions, thanks to the removal of saturated regions.
Still, for analog circuital implementation, we must ensure that $\sigma$ is sufficiently \quotes{simple}. Thus, in the following, we seek an affine separation function, i.e., $\sigma(p) = a^\top p + b$, and we present the following theorem, concerning its existence and computation:
\begin{theorem}[Affine separation] \label{th:reg_sep}
    Let $\mathcal{S}_1$ and $\mathcal{S}_2$ be the unions of two sets of polytopes, such that $\mathcal{S}_1 \cap \mathcal{S}_2 = \emptyset$.
    Let $V_1$ and $V_2$ be the sets of vertices of $\mathcal{S}_1$ and $\mathcal{S}_2$, respectively. Finally, let $\sigma(x) = a^\top x + b$ be an affine function. Then, $\sigma$ separates $\mathcal{S}_1$ and $\mathcal{S}_2$, i.e.,
    \begin{align} \label{eq:th_sep_1}
         \sigma(x) < 0, \;\; \forall\, x \in \mathcal{S}_1, \quad \sigma(x) > 0, \;\; \forall\, x \in \mathcal{S}_2,
    \end{align}
    if and only if the following linear program (LP) is feasible:
    \begin{subequations} \label{eq:reg_sep_op}
        \begin{align}
            \max_{a,b,\varepsilon} \;\;\; & \varepsilon, \\
            \text{\rm s.t.} \;\;\; & \varepsilon > 0,~~\norm{a}_{\infty} \leq 1\\
            & a^\top v^{(1)} + b \leq - \varepsilon, \quad \forall\, v^{(1)} \in V_1, \label{eq:reg_sep_op_a} \\
            & a^\top v^{(2)} + b \geq \phantom{+}\varepsilon, \quad \forall\, v^{(2)} \in V_2. \label{eq:reg_sep_op_b}
    \end{align}
    \end{subequations}
    Moreover, the separation function is given by $\sigma(x) = a^{* \top}x + b^*$, where $(a^*, b^*, \varepsilon^*)$ is the global optimum of problem~\eqref{eq:reg_sep_op}.    
\end{theorem}
\begin{proof}
    ($\Rightarrow$) Since $V_1 \subset \mathcal{S}_1$ and $V_2 \subset \mathcal{S}_2$, Eq.~\eqref{eq:th_sep_1} directly implies the feasibility condition of problem~\eqref{eq:reg_sep_op}. \\
    \noindent
    ($\Leftarrow$) For all positive $\varepsilon$, feasibility of $(a,b)$ implies
    \begin{align} \label{eq:th_sep_2}
        \sigma(v) < 0, \;\; \forall\, v \in V_1, \quad \sigma(v) > 0, \;\; \forall\, v \in V_2.
    \end{align}
    Next, we drop the subscripts $1$ and $2$ to consider both polytopes at the same time. Consider $\lambda \in \mathbb{R}^{|V|}$ such that $\lambda \geq 0$ and $\sum_{i=1}^{|V|} \lambda_i = 1$. Then, by Eq.~\eqref{eq:th_sep_2}, we have that
    \begin{subequations} \label{eq:th_sep_proof_1}
        \begin{align}
            &\lambda_i (a^\top v_i + b) \gtreqless 0, \quad \forall\, i \in \{1,\ldots,|V|\} \label{eq:th_sep_proof_1a} \\
            &a^\top \sum\nolimits_{i=1}^{|V|} \lambda_i v_i + b \sum\nolimits_{i=1}^{|V|} \lambda_i \gtrless 0 \label{eq:th_sep_proof_1b} \\
            &a^\top x + b \gtrless 0, \label{eq:th_sep_proof_1c}
        \end{align}
    \end{subequations}
    where $x = \sum_{i=1}^{|V|} \lambda_i v_i \in \conv(V)$, by definition of convex hull. In Eq.~\eqref{eq:th_sep_proof_1b}, we drop the equality because at least one term of the sum is surely not null by construction. Since $\mathcal{S} \subseteq \conv(V)$, it holds that $x \in \mathcal{S}$, yielding Eq.~\eqref{eq:th_sep_1}.
\end{proof}

\begin{remark}
    The LP feasibility problem~\eqref{eq:reg_sep_op} provides the maximal separation margin $\varepsilon^* > 0$ between the two regions $\mathcal{S}_1$ and $\mathcal{S}_2$, and the separation hyperplane $\sigma$, thereby enhancing the robustness to manufacturing tolerances of the components related to the implementation of $\sigma$.
    The variable $\epsilon$ in the LP~\eqref{eq:reg_sep_op} acts as a tolerance margin for the separating condition. In general, the EMPC formulation~\eqref{eq:empc_policy_sat_sep} may admit infinite affine separation functions. Solving~\eqref{eq:reg_sep_op} enhances the robustness to component tolerances of the subsequent analog implementation by looking for the separation function that maximizes the distance between the separation hyperplane and the sets $\mathcal{S}_1$ and $\mathcal{S}_2$. 
\end{remark}

\begin{remark}
    If the LP problem~\eqref{eq:reg_sep_op} does not admit a feasible solution, then there exists no affine function able to separate $\mathcal{S}_1$ and $\mathcal{S}_2$. In such cases, it is possible to resort to more complex separating functions, e.g., polynomial ones~\cite{paper-kvasnica-2013}, whose analog circuital implementation will be the subject of future research.
    Nevertheless, for the application considered in this paper, we could find an affine separator for the EMPC policy~\eqref{eq:empc_policy_sat_sep} as  shown in  Sec.~\ref{sec:buck_applic}. 
\end{remark}

Hyperplane separation of saturated regions is most effective when a significant proportion of the regions are saturated. Tighter constraints on the input and a smaller magnitude of the input weighting matrix $R$ result in more regions becoming saturated~\cite{paper-grieder-2003}, thus enhancing the effectiveness of this method.

\subsubsection{Removal of Trivial Inequalities}

The EMPC policy~\eqref{eq:mpc_policy} is defined over the polytopic domain $\mathcal{P}$, by Theorem~\ref{th:empc}.
The set $\mathcal{P}$ is typically specified by the user, based on prior knowledge of the admissible values that $p_k$ can take. Thus, any parameter $p_k$ considered in practice always belongs to $\mathcal{P}$.
As a result, in the EMPC policy, the inequalities associated with the facets of $\mathcal{P}$ are trivial (always satisfied) and, therefore, can be removed.

The four reduction techniques are applied sequentially from the highest to lowest abstraction level. Move blocking modifies the optimization problem, defining the initial region partition. Optimal merging and hyperplane separation then reduce the number of regions while preserving the MPC formulation. Finally, removing trivial inequalities refines the geometric description of the remaining regions at the lowest level.

The proposed procedure is key to an efficient EMPC hardware implementation. Since each region and inequality map to circuit components, the reduction strategies directly reduce the required hardware complexity, as detailed below.

\subsection{Circuital Implementation} \label{sec:circuit}
This section presents the circuit implementation of the complexity-reduced EMPC policy, using only commercially available low-latency analog components.

We start by introducing an alternative description of the EMPC policy.
Without loss of generality, let the unsaturated regions be indexed first, i.e., $I_\mathrm{unsat} = \{1,\ldots,N_\mathrm{unsat}\}$, with $N_\mathrm{unsat} \doteq |I_\mathrm{unsat}|$, and let them be described as
\begin{equation}
    \mathcal{R}_j = \{p_k \in \mathbb{R}^{n_p} : H_j p_k \le h_j\},\quad j = 1,\ldots,N_\mathrm{unsat},
\end{equation}
where $H_j \in \mathbb{R}^{N_j \times n_p}$, $h_j \in \mathbb{R}^{N_j}$, and $N_j$ is the number of scalar inequalities defining region $\mathcal{R}_j$. Accordingly, we define the Boolean variables associated with each unsaturated region as
\begin{equation} \label{eq:cmp_bool_vars1}
    r_j =
    \begin{cases}
    1 & \text{if } H_j p_k \le h_j,\\
    0 & \text{otherwise},
    \end{cases}
    \quad j=1,\dots,N_\mathrm{unsat}.
\end{equation}
Given these, we define variables $\beta_i \in \{0, 1\}$ as \begin{equation}\label{eq:law_indicator}
    \beta_i = \bigoplus\nolimits_{j \in \mathcal{I}_i} r_j, \quad i \in I_\mathrm{unsat}^{\mathrm{unique}}
\end{equation}
that assumes logical value $1$ whenever the affine law $K_i p_k + l_i$ must be selected. In Eq.~\eqref{eq:law_indicator}, $\mathcal{I}_i \subseteq I_\mathrm{unsat}$ is the set of region indices sharing the law $K_i p_k + l_i$ and $I_\mathrm{unsat}^{\mathrm{unique}} \subseteq I_\mathrm{unsat}$ is the set of indices of unsaturated regions that do not share the same law. We denote by $N \doteq |I_\mathrm{unsat}^{\mathrm{unique}}|$ the number of unique unsaturated affine functions. Finally, by defining the saturation signals
\begin{equation}\label{eq:empc_policy_binary_b}
    s =
\begin{cases}
1 & \text{if } a^{*\top} p_k + b^* > 0,\\
0 & \text{otherwise},
\end{cases}
\quad
s_a = \bigodot_{i\in I_\mathrm{unsat}^{\mathrm{unique}}} \overline{\beta_i},
\end{equation}
the EMPC policy~\eqref{eq:empc_policy_sat_sep} is equivalent to
\begin{equation}\label{eq:empc_policy_final}
u_k = \pi(p_k) =
\begin{cases}
K_i p_k + l_i & \text{if } \beta_i,\;\; i \in {I}_\mathrm{unsat}^\mathrm{unique},\\
u_{\mathrm{ub}} & \text{if } s_a \odot s,\\
u_{\mathrm{lb}} & \text{if } s_a \odot \overline{s}.
\end{cases}
\end{equation}

\begin{remark}
    The construction of the signals $\beta_i$ in Eq.~\eqref{eq:law_indicator} takes into account the non-disjoint optimal merging strategy, described in Sec.~\ref{sec:explicit_mpc}. Specifically, in the case of an overlap between regions sharing the same $i$-th affine function, all the variables $r_j$ with $j \in \mathcal{I}_i$ are combined using an OR operator so that the $i$-th affine function is selected if $p_k$ belongs to $\bigcup_{j \in \mathcal{I}_{i}} \mathcal{R}_j$, i.e., the whole domain of the shared affine function.
\end{remark}

\begin{figure}[t!]
    \centering
    \includegraphics[width=0.9\linewidth]{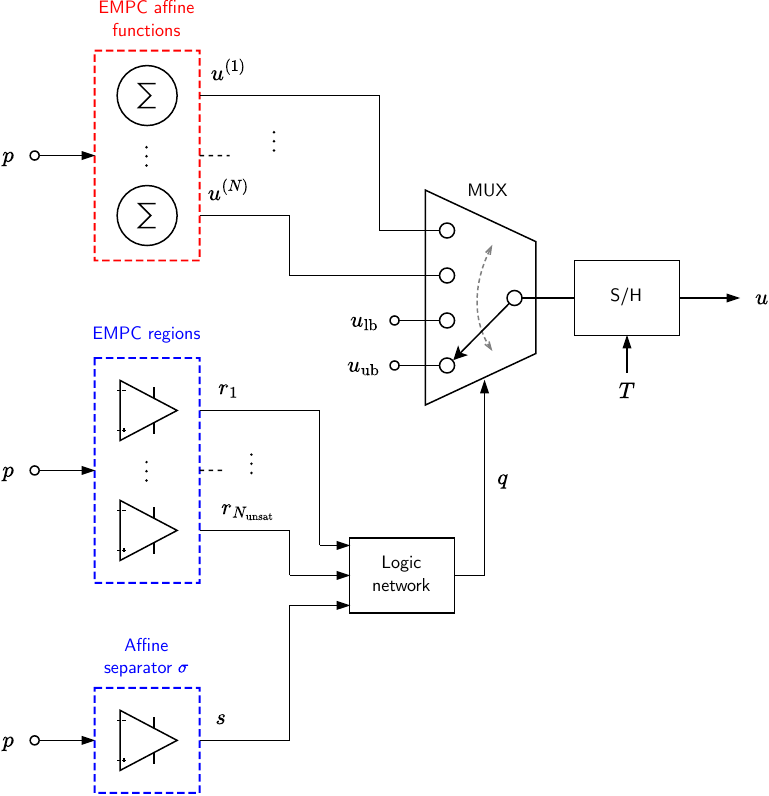}
    \caption{EMPC implementation: high-level schematic.}
    \vspace{-\baselineskip}
    \label{fig:implem_mpc_top}
\end{figure}

We implement the EMPC policy~\eqref{eq:empc_policy_final} using: (i) one multiplexer (MUX), (ii) a set of generalized adders, (iii) a set of comparators, and (iv) a small logic gate network, according to the circuit depicted in Fig.~\ref{fig:implem_mpc_top}. A sample and hold (S/H) circuit is included after the MUX to ensure proper sampling and avoid inter-sample oscillations of the control input.

Below, we detail every stage of the design for the four main building blocks of the EMPC circuit. 

\subsubsection{Multiplexer}

It is the main component in our EMPC circuit design.
We use an analog multiplexer to implement the by-case PWA control policy in Eq.~\eqref{eq:empc_policy_final}. Specifically, we use a multiplexer with $N+2$ inputs: one for each unique unsaturated affine function, plus the two saturated input values $u_\mathrm{ub}$ and $u_\mathrm{lb}$. While $u_\mathrm{ub}$ and $u_\mathrm{lb}$ are trivially imposed through a constant voltage, the inputs corresponding to the affine functions need to be computed online, given the current value of states and disturbances. To this end, we introduce generalized adders.

\subsubsection{Generalized Adders}

Generalized adders are a standard circuit in analog electronics~\cite{sedra2020microelectronic} and their design is easily automated, as detailed in the following.
We employ generalized adders to generate the control input $u^{(i)}$ given by each $i$-th unique unsaturated affine function in the EMPC policy~\eqref{eq:empc_policy_final},
\begin{align} \label{eq:buck_empc_ga_aff_fun}
    u^{(i)} = K_i p + l_i, \quad i \in I_\mathrm{unsat}^{\mathrm{unique}},
\end{align}
where, for notational convenience, we omit the explicit time dependence on $k$. 

Without loss of generality, we consider the case $n_u = 1$ (i.e., $K_i \in \mathbb{R}^{1 \times n_p}$ and $l_i \in \mathbb{R}$). In this setting, being $u^{(i)}$ a scalar, each affine function requires a single generalized adder, so $N$ in total. If $n_u > 1$, then a generalized adder is needed for each component $(u^{(i)})_j$, $j=1,\ldots,n_u$, resulting in $n_u$ generalized adders for each affine function, so $n_u N$ in total.

For each generalized adder, the circuit inputs are the voltage measurements of the EMPC parameters $p$ and a positive constant voltage $V_0$ to implement the constant offset $l_i$, which can be arbitrarily chosen; the circuit output is a voltage equal to $u^{(i)}$.
Let us append $V_0$ to $p$, i.e., $p_{n_p+1} = V_0$. Also, let $K_{i,j} = (K_i)_j$ and append $l_i$ to $K_i$, i.e., $K_{i,n_p+1} = l_i/V_0$. Then, we can equivalently rewrite Eq.~\eqref{eq:buck_empc_ga_aff_fun} as follows:
%
    \begin{align}\label{eq:buck_empc_gen_sum}
        u^{(i)} = \; & \sum_{j=1}^{n_p+1} K_{i,j} \, p_j = \sum_{k=1}^{M_i} g_{i,k}^{(+)} \, p_{i,k}^{(+)} - \sum_{k=1}^{N_i} g_{i,k}^{(-)} \, p_{i,k}^{(-)}, 
    \end{align}
%
where, letting $M_i,N_i \in \mathbb{N}$ the number of inputs with positive and negative gains, respectively, we defined $g_{i,k}^{(+)}=K_{i,j}\geq 0$ for $k=1,\dots,M_i$, $-g_{i,k}^{(-)}=K_{i,j}< 0$ for $k=1,\dots,N_i$, and $p_{i,k}^{(+)}$, $p_{i,k}^{(-)}$ the inputs associated with gains $g_{i,k}^{(+)}$ and $g_{i,k}^{(-)}$, respectively. In the following, we omit the adder index $i$, except in $N_i$ and $M_i$.

\begin{figure}[t!]
    \centering
    \includegraphics[width=0.8\linewidth]{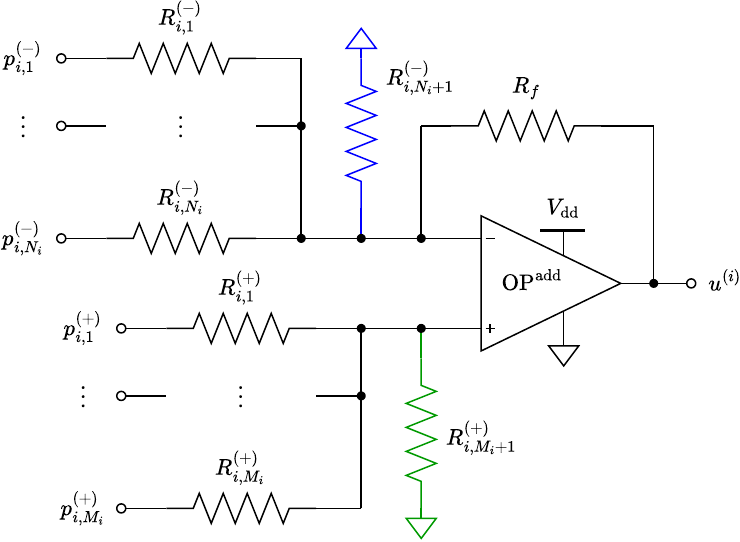}
    \caption{Generalized adder for the $i$-th affine function. The blue resistor is used if $K_i^{(+)}\geq K_i^{(-)}+1$, while the green one is used otherwise.}
    \vspace{-\baselineskip}
    \label{fig:genadd}
\end{figure}

Let us define the total positive and negative gains, i.e.,
\begin{equation} \label{eq:buck_empc_ga_tot_gains}
    K^{(+)}=\sum\nolimits_{k=1}^{M_i} g_{k}^{(+)}, \quad K^{(-)}=\sum\nolimits_{k=1}^{N_i} g_{k}^{(-)}.
\end{equation}
We identify two topologically distinct circuits for the considered design, depending on whether $K^{(+)} \geq K^{(-)} + 1$ or $K^{(+)} < K^{(-)} + 1$. Both topologies are reported in Fig.~\ref{fig:genadd}.

In the first case, with $K^{(+)} \geq K^{(-)} + 1$, the circuit includes the additional resistance $R_{N_i+1}^{(-)}$, connected to the inverting input of the OP-AMP $\mathrm{OP}^\mathrm{add}$, as shown in Fig.~\ref{fig:genadd} (blue resistor) to increase the overall positive gain. In this case, if we ensure
\begin{equation} \label{eq:buck_empc_ga_cond_kpgrkn}
    G_f + \sum\nolimits_{k=1}^{N_i+1}G_{k}^{(-)} = \sum\nolimits_{k=1}^{M_i}G_{k}^{(+)}, 
\end{equation}
where $G_f=1/R_f$, $G_{k}^{(-)}=1/R_{k}^{(-)}$, and $G_{k}^{(+)}=1/R_{k}^{(+)}$, we obtain the following:
\begin{equation} \label{eq:buck_empc_ga_gains_kpgrkn}
    g_{k}^{(-)} = G_{k}^{(-)}G_f^{-1}, \quad g_{k}^{(+)} = G_{k}^{(+)}G_f^{-1}.
\end{equation}
Therefore, fixed a value for $R_f$, we can trivially obtain $R_k^{(-)}$ and $R_k^{(+)}$ from Eq.~\eqref{eq:buck_empc_ga_gains_kpgrkn} and, next, $R_{N_i+1}^{(-)}$ from Eq.~\eqref{eq:buck_empc_ga_cond_kpgrkn}.

Similarly, in the second case, with $K^{(+)}<K^{(-)}+1$, we reduce the overall positive gain by including the additional resistance $R_{M_i+1}^{(+)}$ to the positive input of $\mathrm{OP}^\mathrm{add}$, as shown in Fig.~\ref{fig:genadd} (green resistor). In this case, if
\begin{equation}\label{eq:buck_empc_ga_cond_kngrkp}
    G_f + \sum\nolimits_{k=1}^{N_i}G_{k}^{(-)} = \sum\nolimits_{k=1}^{M_i+1}G_{k}^{(+)}, 
\end{equation}
we obtain, as for the previous case, the design equations in Eq.~\eqref{eq:buck_empc_ga_gains_kpgrkn}, through which we can compute $R_k^{(-)}$ and $R_k^{(+)}$. Finally, $R_{M_i+1}^{(+)}$ is computed using Eq.~\eqref{eq:buck_empc_ga_cond_kngrkp}.

Adders constitute the most expensive part of the device, as they require OP-AMPs. Still, the overall number of required adders is $N$, i.e., the number of unique unsaturated affine functions of the EMPC policy~\eqref{eq:empc_policy_final}, which can be reduced through the techniques described in Sec.~\ref{sec:explicit_mpc}.

\subsubsection{Comparators}

We employ a set of comparators to obtain the Boolean signals $(r_j)_{j =1}^{N_\mathrm{unsat}}$ and $s$, as defined in Eqs.~\eqref{eq:cmp_bool_vars1}--\eqref{eq:empc_policy_binary_b}.
For each $r_j$, let us define the auxiliary Boolean signal
\begin{align}
    r_{j,k} = \begin{cases}
        1 & \textrm{if} \;\; -(H_j)_{k} \, p + (h_j)_{k} \geq 0, \\
        0 & \textrm{otherwise},
    \end{cases}
\end{align}
where $(H_j)_{k}$ and $(h_j)_{k}$ are the $k$-th row of matrix $H_j$ and the $k$-th element of vector $h_j$, respectively.
Then, we have that $r_j = \bigodot_{k=1}^{N_j} r_{j,k}$, $j = 1, \ldots, N_\mathrm{unsat}$, which can be easily realized through logic AND gates.

\begin{figure}[t!]
    \centering
    \includegraphics[width=0.8\linewidth]{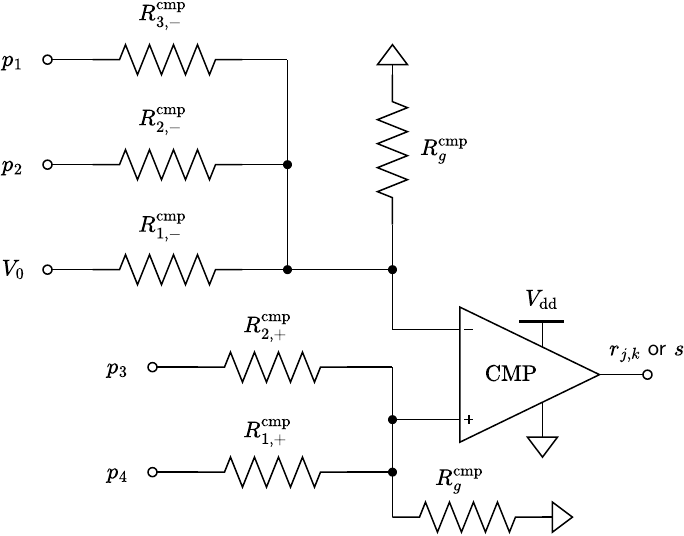}
    \caption{Comparator (case with $n_p = 4$, $\vert{I_p}\vert=2$, and $\vert I_m\vert=3$).}
    \label{fig:cmp}
\end{figure}

Obtaining each of the Boolean signals $r_{j,k}$ and $s$ requires evaluating a set of inequalities of the kind
\begin{align} \label{eq:buck_empc_sum_cmp}
    [\alpha_1, \; \ldots, \; \alpha_{n_p}] \, p + \alpha_{n_p+1} V_0 \geq 0,
\end{align}
where $V_0$ is a positive constant voltage, which can be arbitrarily chosen (it is also employed in the generalized adders), and $(\alpha_i)_{i=1}^{n_p+1}$ are gains depending on $(H_j)_k$, $(h_j)_k$, $a^*$, and $b^*$, whose sign is not known a-priori.
Let us append $V_0$ to $p$, i.e., $p_{n_p+1} = V_0$.
We realize the comparison altogether with the sums in Eq.~\eqref{eq:buck_empc_sum_cmp} through the circuit topology in Fig.~\ref{fig:cmp}. The selected circuit compares the voltage on the \quotes{$+$} terminal with that on the \quotes{$-$} terminal; these voltages are obtained through voltage dividers.
To determine the resistance values, we rewrite Eq.~\eqref{eq:buck_empc_sum_cmp} as
\begin{align}
    \sum\nolimits_{i \in I_p} \gamma_i \, p_i \geq \sum\nolimits_{i \in I_m} \gamma_i \, p_i,
\end{align}
where $I_p, \, I_m \subseteq \{1, \ldots, n_p+1\}$ are the set of indices for which $\alpha_i > 0$ and $\alpha_i \leq 0$ in Eq.~\eqref{eq:buck_empc_sum_cmp}, respectively, and
\begin{align}
    \gamma_i = \frac{c \, |\alpha_i|}{\max_{j \in \{1, \ldots, n_p+1\}} |\alpha_j|}, \quad c \in (0,1)
\end{align}
are normalized and rescaled versions of the gains $\alpha_i$.
In this way, the gains $\gamma_i$ associated with each $p_i$ are such that $0 < \gamma_i < 1$, for all $i=1,\ldots,n_p+1$. This enables the use of inexpensive voltage dividers for comparisons, eliminating the need for additional OP-AMPs.

Consider the networks of $|I_p|+1$ and $|I_m|+1$ resistors on the \quotes{$+$} and \quotes{$-$} terminal of the comparator, respectively.
Let $R_g^\mathrm{cmp}$ be the resistor connected to ground, $R_i^\mathrm{cmp}$ the resistor associated with the input $p_i$, $G_g^\mathrm{cmp} = 1/R_g^\mathrm{cmp}$ and $G_i^\mathrm{cmp} = 1/R_i^\mathrm{cmp}$ the corresponding conductances.
Then, the voltages on the two terminals are
\begin{align}
    v_+ = \sum_{i\in I_p} \frac{ (G_g^\mathrm{cmp}+\sum_{j \neq i} G_j^\mathrm{cmp})^{-1} }{ R_i^\mathrm{cmp} + (G_g^\mathrm{cmp}+\sum_{j \neq i} G_j^\mathrm{cmp})^{-1}} p_i
\end{align}
and similarly for $v_-$, with the sum over $i \in I_m$.

Imposing the gains match, i.e.,
\begin{align}
    \gamma_i = \frac{ (G_g^\mathrm{cmp}+\sum_{j \neq i} G_j^\mathrm{cmp})^{-1} }{ R_i^\mathrm{cmp} + (G_g^\mathrm{cmp}+\sum_{j \neq i} G_j^\mathrm{cmp})^{-1}}, \quad i \in I_p,
\end{align}
is equivalent to
\begin{align} \label{eq:buck_empc_cmp_resist}
    \gamma_i \Big(G_g^\mathrm{cmp}+\sum\nolimits_{j \in I_p}G_j^\mathrm{cmp}\Big) = G_i^\mathrm{cmp}, \quad i \in I_p.
\end{align}
For any fixed resistor $R_g^\mathrm{cmp}$, Eq.~\eqref{eq:buck_empc_cmp_resist} is a linear system of $|I_p|$ equations in $|I_p|$ unknowns, whose matrix form is
\begin{equation} \label{eq:buck_empc_cmp_resist_sys}
    \begin{bmatrix}
        \gamma_1-1 & \gamma_1 & \dots & \gamma_1 \\
        \gamma_2 & \gamma_2-1 & \dots & \gamma_2\\
        \vdots & \vdots & \ddots & \vdots \\
        \gamma_{|I_p|} & \gamma_{|I_p|} & \dots & \gamma_{|I_p|}-1
    \end{bmatrix} \!\!\! \begin{bmatrix}
        G_1^\mathrm{cmp} \\ G_2^\mathrm{cmp} \\ \vdots \\ G_{|I_p|}^\mathrm{cmp} 
    \end{bmatrix} \! = \! - G_g^\mathrm{cmp} \begin{bmatrix}
        \gamma_1 \\ \gamma_2 \\ \vdots  \\ \gamma_{|I_p|} 
    \end{bmatrix} \! .
\end{equation}
The matrix in Eq.~\eqref{eq:buck_empc_cmp_resist_sys} is always full-rank, thus admitting a unique solution for the conductances $G_i^\mathrm{cmp}$. Then, resistors $R_i^\mathrm{cmp}$ are easily obtained by taking the inverse. The very same procedure applies to the network at the \quotes{$-$} terminal.

Overall, the number of comparators is upper-bounded by $1+\sum_{j=1}^{N_\mathrm{unsat}} N_j$, which is often effectively optimized through the complexity-reduction techniques described in Sec.~\ref{sec:explicit_mpc}, particularly the removal of trivial inequalities and the non-disjoint optimal merging strategy, which reduces both the number of EMPC regions and the number of facets.
This number can be further reduced if different regions share the same inequality or use opposite ones (i.e., two or more regions have facets on the same hyperplane); in such scenarios, we may reuse comparators from other regions, eventually introducing a NOT logic gate.

\subsubsection{Logic Network} \label{sec:circuit_logic_net}
A network of logic gates is needed to drive the selection signal $q$ of the MUX, using the Boolean signals $(r_j)_{j=1}^{N_\mathrm{unsat}}$ and $s$, according to Eqs.~\eqref{eq:law_indicator}--\eqref{eq:empc_policy_final}.
Specifically, the signal $q = (q_i)_{i=1}^{\lceil\log_2(N+2)\rceil}$ is binary encoded to select each unsaturated affine function and the two saturated values $u_\mathrm{lb}$ and $u_\mathrm{ub}$.
The logic network implementing $q$ can be easily designed using standard logic function optimization methods, such as Karnaugh maps.

\subsection{Comparison with the literature}
\label{sec:comp_circuits}
The proposed approach presents several structural differences with respect to existing MPC implementations. Compared to digital solutions (such as~\cite{cim25,mariethoz_comparison_2010} based on microcontrollers, or~\cite{WillsTCST12_fpga,liu_fast_2018} based on FPGA), it avoids the need for ADC and DAC interfaces, thereby reducing conversion-induced delays and hardware overhead. With respect to the prior analog implementation proposed in~\cite{vichik_solving_2014}, all the circuit OP-AMPs operate in their linear region, thereby avoiding slew-rate limitations and enabling lower latency and higher achievable switching frequencies.

In addition, the proposed design links circuit complexity to the number of unsaturated regions in the EMPC law, which can be influenced through MPC parameters' choices. For instance, a shorter horizon, fewer constraints, and the use of move blocking directly reduce the number of regions as shown in Eq.~\eqref{eq:reg_bnd_binomial}, and the use of input regularization enhances the effectiveness of the hyperplane separation of saturated regions. This provides some flexibility in balancing closed-loop performance and implementation complexity. In contrast, in~\cite{vichik_solving_2014} the number of required OP-AMPs is directly tied to the number of constraints and, thus, cannot benefit from any of the proposed simplification strategies, directly limiting the overall design flexibility. Furthermore, in our design, fewer OP-AMPs are needed: we only require $1$ OP-AMP per region, versus the $2$ OP-AMPs per inequality in~\cite{vichik_solving_2014}.

\section{Application to DC-DC Buck Converters} \label{sec:buck_applic}

In this section, we apply the proposed approach to design a fully analog MPC-controlled DC-DC Buck converter. 

\subsection{Buck Converter Mathematical Model} \label{sec:buck_model}
\begin{figure}[t!]
    \centering
    \includegraphics[width=\linewidth]{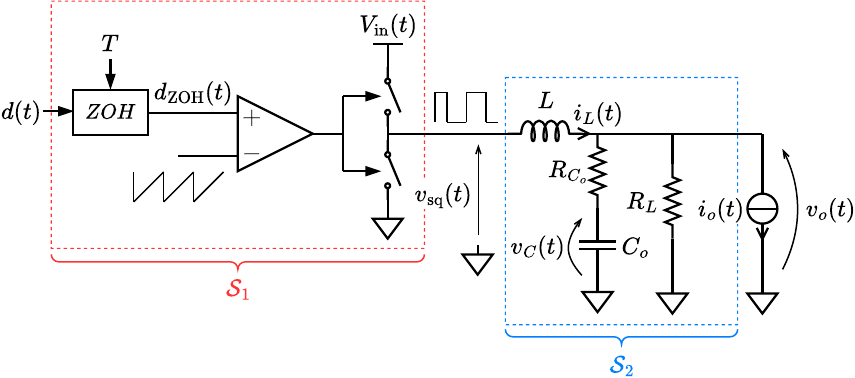}
    \caption{Buck converter circuit diagram.}
    \vspace{-\baselineskip}
    \label{fig:buck_circuit_diagram}
\end{figure}

We consider a Buck converter operating at a fixed switching frequency $f_\mathrm{sw} = \frac{1}{T}$.
With reference to the circuit diagram of the converter depicted in Fig.~\ref{fig:buck_circuit_diagram}, we denote as $i_L(t)$ the inductor current, $v_C(t)$ the capacitor voltage, $d(t)$ the duty cycle of the square wave voltage $v_\mathrm{sq}(t)$, $i_o(t)$ the drawn output current, $V_\mathrm{in}(t)$ the supply voltage, and $v_o(t)$ the output voltage. We denote as $\overline{V}_\mathrm{in}$ and $v_\mathrm{in}(t)$ the nominal value and variation of $V_\mathrm{in}(t)$, i.e., $V_\mathrm{in}(t) = \overline{V}_\mathrm{in} + v_\mathrm{in}(t)$. The main control task is to regulate $v_o(t)$ at a reference $\overline{V}_o$.

We rely on the first principles of physics to relate the above quantities and derive the mathematical model of the Buck converter.
Specifically, we describe the Buck converter as the cascade of two subsystems (see Fig.~\ref{fig:buck_circuit_diagram}).

The first subsystem, $\mathcal{S}_1$, takes as inputs the duty cycle $d(t)$ in the form of a voltage signal ranging in $[0,1]\;\unit{V}$, and the supply voltage $V_\mathrm{in}(t)$. The output of $\mathcal{S}_1$ is the voltage $v_\mathrm{sq}(t)$. This subsystem consists of three parts: a zero-order hold (ZOH), a comparator, and two switches.
The ZOH samples $d(t)$ at each time instant $t=kT$, $k \in \mathbb{Z}_{\geq 0}$, and holds it constant at the value $d_k = d(kT)$ over the $k$-th switching period $I_k = [kT, (k+1)T)$. The ZOH output signal is given by $d_\mathrm{ZOH}(t) = d_k$, for all $t \in I_k$.
Then, $d_\mathrm{ZOH}(t)$ is compared to a sawtooth wave, with fixed frequency $f_\mathrm{sw}$ and ranging in $[0,1]\;\unit{V}$. This operation is performed by means of a comparator. When the sawtooth value is below $d_\mathrm{ZOH}(t)$, the comparator outputs a voltage $V_H$; conversely, when the sawtooth value is larger than $d_\mathrm{ZOH}(t)$, the comparator outputs a voltage $V_L$.
The comparator output signal drives the two switches driven in anti-synchronized mode. One switch, called the high side, is connected to $V_\mathrm{in}(t)$ and is open when the comparator outputs $V_H$; the other, called the low side, is connected to the ground and is open if the comparator outputs $V_L$. As a result, the subsystem $\mathcal{S}_1$ outputs a square wave $v_\mathrm{sq}(t)$ with duty cycle $d(t)$ and amplitude $V_\mathrm{in}(t)$, according to:
\begin{align} \label{eq:def_S1_block}
    \hspace{-5pt} \mathcal{S}_1(d(t),V_\mathrm{in}(t)) \doteq \begin{cases}
        V_\mathrm{in}(t), & \!\! t \in [kT, (k+d_k)T), \\
        0, & \!\! t \in [(k+d_k)T, (k+1)T).
    \end{cases}
\end{align}

The subsystem $\mathcal{S}_2$ is dynamical, linear and time-invariant (LTI). Its inputs are the square wave voltage $v_\mathrm{sq}(t)$ from $\mathcal{S}_1$ and the output current $i_o(t)$. The output of $\mathcal{S}_2$ is the controlled output voltage $v_o(t)$.
Following the standard approach in circuit theory, we define the state vector as $x(t) = [i_L(t), \; v_C(t)]^\top$.
With this choice, we obtain the following state-space model for $\mathcal{S}_2$:
\begin{align} \label{eq:sys_ct_buck}
    \mathcal{S}_2: \left\{
    \begin{aligned}
        \dot x(t) &= A_c x(t) + B_{c,1} i_o(t) + B_{c,2} v_\mathrm{sq}(t), \\
        v_o(t) &= C_c x(t) + D_1 i_o(t),
    \end{aligned} \right.
\end{align}
where
\begin{subequations} \label{eq:sys_ct_buck_mat}
\begin{align}
    & \hspace{-10pt} A_c = 
    \begin{bmatrix}
        -\frac{R_L \parallel R_{C_o}}{L} & -\frac{R_L}{L(R_{C_o} + R_L)} \\
        \frac{R_L}{C_o(R_{C_o} + R_L)} &  -\frac{1}{C_o(R_{C_o} + R_L)}
    \end{bmatrix},  \\
    & \hspace{-10pt} B_{c,1} = \begin{bmatrix}
        \frac{R_L \parallel R_{C_o}}{L} \\
        -\frac{R_L}{C_o(R_{C_o} + R_L)}
    \end{bmatrix}, \quad  B_{c,2} = \begin{bmatrix} \frac{1}{L} \\ 0 \end{bmatrix}, \\
    & \hspace{-10pt} C_c = \begin{bmatrix}
        R_L \parallel R_{C_o} & \frac{R_L}{R_L + R_{C_o}}
    \end{bmatrix}, \quad
    D_{c,1} = - R_L \parallel R_{C_o}.
\end{align}
\end{subequations}
We refer the reader to~\cite{book-erickson-2020} for details.

Overall, the model of the entire plant is a nonlinear system described by the interconnection of $\mathcal{S}_1$ and $\mathcal{S}_2$:
%
    \begin{align} \label{eq:sys_ct_buck_complete}
        v_o(t) = \mathcal{S}_2(i_o(t), \mathcal{S}_1(d(t),V_\mathrm{in}(t))).
    \end{align}
%
Notice that $v_o(t)$ depends on the $d(t)$, which we use as control input, but also on $i_o(t)$ and $\overline{V}_\mathrm{in}$. The latter two quantities are external signals that we can not act upon; therefore, we shall consider their variations as disturbances to be compensated, i.e., $\nu_1(t) = i_o(t)$ and $\nu_2(t) = V_\mathrm{in}(t) - \overline{V}_\mathrm{in} = v_\mathrm{in}(t)$. 

According to the MPC formulation in Sec.~\ref{sec:probl_formul}, we need to describe the plant under control in Eq.~\eqref{eq:sys_ct_buck_complete} as a DT affine system, matching Eq.~\eqref{eq:sys_dt_lti}. To this end, we perform discretization and linearization of the model~\eqref{eq:sys_ct_buck_complete}. 

\subsubsection{Model Discretization}

We select the switching period $T$ as the discrete time step for discretization. Henceforth, any quantity sampled at the switching time instants $t = kT$ will be denoted using the subscript $k$, i.e., $\star_k = \star(kT)$.

We address discretization by integrating Eq.~\eqref{eq:sys_ct_buck} over the $k$-th switching period $I_k$. Thanks to the linearity of Eq.~\eqref{eq:sys_ct_buck}, we can evaluate the model prediction $x_{k+1}$ in closed form using the following convolution integral:
\begin{align} \label{eq:conv_int_1}
    & x_{k+1} = e^{A_c T}x_k \, + \nonumber \\
    & \int_{kT}^{(k+1)T} e^{A_c((k+1)T-\tau)} (B_{c,1} i_o(t) + B_{c,2} v_\mathrm{sq}(t)) d\tau.
\end{align}
By linearity, Eq.~\eqref{eq:conv_int_1} is equivalent to
\begin{align} \label{eq:conv_int_2}
    x_{k+1} = \; & e^{A_c T}x_k + e^{A_c(k+1)T}\int_{kT}^{(k+1)T} e^{-A_c\tau} B_{c,1} i_{o}(t) d\tau \nonumber \\
    & + e^{A_c(k+1)T}\int_{kT}^{(k+1)T} e^{-A_c\tau} B_{c,2} v_\mathrm{sq}(t) d\tau.
\end{align}
The integrals in Eq.~\eqref{eq:conv_int_2} are solved in closed form. To this aim, we introduce the following assumption:

\begin{assumption} \label{ass:piecewise_const_dist}
    The input voltage $V_\mathrm{in}(t)$ and the output current disturbance $i_{o}(t)$ remain constant over the switching periods, i.e., $V_\mathrm{in}(t) = V_{\mathrm{in},k}$ and $i_o(t) = i_{o,k}$ for all $t \in I_k$ and $k \geq 0$.
\end{assumption}

\begin{remark}
    Note that, Assumption~\ref{ass:piecewise_const_dist} is respected with a good approximation as long as: i) the input voltage $V_\mathrm{in}(t)$ slowly varies (i.e., its bandwidth is much lower than $f_\mathrm{sw}$); ii) the output current $i_o(t)$ is constant almost everywhere (i.e., it is well modeled by a piecewise-constant signal, since it only changes due to sudden load connections or removals).
\end{remark}

Under Assumption~\ref{ass:piecewise_const_dist} and using $v_\mathrm{sq}(t) = \mathcal{S}_1(d(t),V_\mathrm{in}(t))$ as defined in Eq.~\eqref{eq:def_S1_block}, Eq.~\eqref{eq:conv_int_2} simplifies to
\begin{align} \label{eq:conv_int_3}
    x_{k+1} = \; & e^{A_c T}x_k + e^{A_c(k+1)T}\int_{kT}^{(k+1)T} e^{-A_c\tau} B_{c,1} i_{o,k} d\tau \nonumber \\
    & + e^{A_c(k+1)T}\int_{kT}^{(k+d_k)T} e^{-A_c\tau} B_{c,2} V_{\mathrm{in},k} d\tau.
\end{align}
Solving the integrals in Eq.~\eqref{eq:conv_int_3} leads to the DT model
\begin{align} \label{eq:sys_dt_buck_nl}
    x_{k+1} = \; & f(x_k, d_k, i_{o,k}, V_{\mathrm{in},k}) \nonumber \\
    \doteq \; & e^{A_c T}x_k + (e^{A_c T}-I)A_c^{-1}B_{c,1} i_{o,k} \nonumber \\
    & + e^{A_c T}(I - e^{-A_c d_k T})A_c^{-1} B_{c,2} V_{\mathrm{in},k}.
\end{align}

\begin{proposition} \label{prop:sys_dt_buck_nl_equil}
    Let us consider constant disturbances $i_{o,k} = i_o$ and $v_{\mathrm{in},k} = {v}_{\mathrm{in}}$ for all $k \geq 0$. The system~\eqref{eq:sys_dt_buck_nl} admits a unique equilibrium point satisfying the output regulation task towards the desired constant reference $\overline{V}_{o}$, i.e., there exist unique $\overline{x} \in \mathbb{R}^{n}$ and $\overline{D} \in \mathbb{R}$ such that
    \begin{subequations}
        \begin{align}
            \overline{x} &= f(\overline{x},\overline{D},i_o,
            {V}_\mathrm{in}), \label{eq:equil_nl_cond1} \\
            \overline{V}_{o} &= C \overline{x}. \label{eq:equil_nl_cond2}
        \end{align}
    \end{subequations}
\end{proposition}
\begin{proof}
    Solving Eq.~\eqref{eq:equil_nl_cond1} for $\overline{x}$ and using Eq.~\eqref{eq:equil_nl_cond2} yields
    \begin{equation} \label{eq:equil_explicit}
       \overline{V}_{o} = C (I-A)^{-1} (A-A^{1-\overline{D}}) A_c^{-1} B_{c,2} {V}_\mathrm{in} - \rho_1 i_o,
    \end{equation}
    where $A = e^{A_c T}$ and $\rho_1 = C A_c^{-1} B_{c,1} \approx 0$ is the DC-gain of the linear subsystem $\mathcal{S}_2$ in Eq.~\eqref{eq:sys_ct_buck} for input $i_o$. Existence is guaranteed by noting that the scalar function of $\overline{D}$ on the right-hand side of Eq.~\eqref{eq:equil_explicit}, defined over the interval $0 \leq \overline{D} \leq 1$, admits as image the interval $-\rho_1 i_o \leq \overline{V}_o \leq -\rho_1 i_o + \rho_2 V_\mathrm{in}$, where $\rho_2 = C A_c^{-1} B_{c,2} \approx 1$ is the DC-gain of the linear subsystem $\mathcal{S}_2$ in Eq.~\eqref{eq:sys_ct_buck} for input $V_\mathrm{in}$. Uniqueness is a direct consequence of the fact that the function $\overline{V}_o(\overline{D})$ is monotonically strictly increasing.
\end{proof}
\begin{remark}
    The exact value of $\overline{D}$ can be computed, e.g., using Newton's method to solve the nonlinear equation~\eqref{eq:equil_explicit}. Also, we notice that in the undisturbed case, i.e., $i_{o,k} = 0$ and $V_{\mathrm{in},k} = \overline{V}_{\mathrm{in}}$, taking a linear approximation of the exponential function leads to $\overline{D}=\frac{\overline{V}_o}{\overline{V}_\mathrm{in}}$, which is a good approximation in all our numerical tests and matches the classical assessment of steady-state duty-cycle as established in, e.g.,~\cite{book-erickson-2020}.
\end{remark}

\subsubsection{Model Linearization} We proceed with the linearization of the model in Eq.~\eqref{eq:sys_dt_buck_nl}. Specifically, a nonlinearity in $d_k$ and $V_{\mathrm{in},k}$ arises from the presence of due to the presence of the term $e^{-A_c d_k T} V_{\mathrm{in},k}$. We introduce the following assumption:
\begin{assumption} \label{ass:small_signal}
    The sequences $d_k$, $V_{\mathrm{in},k}$ are well-described by
    \begin{align}
        d_k = \overline{D} + \delta_k, \quad V_{\mathrm{in},k} = \overline{V}_\mathrm{in} + v_{\mathrm{in},k},
    \end{align}
    where $\delta_k$ and $v_{\mathrm{in},k}$ are small variations around the nominal values $\overline{D}$ and $\overline V_{\mathrm{in}}$, respectively. 
\end{assumption}
\begin{remark}
    Assumption~\ref{ass:small_signal}, which is standard in set-point regulation for power converters (see, e.g., \cite[Chapter 1]{Gibbard15}), is only used to construct the DT MPC prediction model~\eqref{eq:mpc_op_b},~\eqref{eq:mpc_op_c}, but not critical for stability; as we can formally guarantee stability of the closed-loop system considering the non-approximated nonlinear model in Eq.~\eqref{eq:sys_dt_buck_nl}; see Sec.~\ref{sec:stability}.
\end{remark}

Let us consider the nonlinearity of model~\eqref{eq:sys_dt_buck_nl}, i.e.,
\begin{align} \label{eq:sys_dt_buck_nl_term}
    g(\delta_k,v_{\mathrm{in},k}) \doteq e^{A_c T} (I - e^{-A_c (\overline{D} + \delta_k) T} ) A_c^{-1} B_{c,2} (\overline{V}_\mathrm{in} + v_{\mathrm{in},k}).
\end{align}
Under Assumption~\ref{ass:small_signal}, we can linearize Eq.~\eqref{eq:sys_dt_buck_nl_term} around the point $(\delta_k,v_{\mathrm{in},k}) = (0,0)$ as follows:
\begin{align} \label{eq:sys_dt_buck_nl_term_relaxed}
     & g(\delta_k,v_{\mathrm{in},k}) \approx \; g(0,0) + \frac{\partial g}{\partial \delta_k}(0,0)\,\delta_k + \frac{\partial g}{\partial v_{\mathrm{in},k}}(0,0)\,v_{\mathrm{in},k} \nonumber \\
     & \quad = e^{A_c(1-\overline{D})T}T B_{c,2} \overline{V}_\mathrm{in} \, \delta_k  \\
    & \quad ~~ + e^{A_c T}(I-e^{-A_c \overline{D} T})A_c^{-1}B_{c,2} \, (\overline{V}_\mathrm{in} + v_{\mathrm{in},k}).\nonumber
\end{align}
Finally, using Eq.~\eqref{eq:sys_dt_buck_nl_term_relaxed} to replace $g(\delta,v_{\mathrm{in},k})$ in Eq.~\eqref{eq:sys_dt_buck_nl}, we obtain a linearization matching Eq.~\eqref{eq:sys_dt_lti}, with $y_k = v_{o,k}$ and
\begin{subequations} \label{eq:sys_dt_buck_mat}
    \begin{align}
        &A = e^{A_c T}, \quad B = e^{A_c (1-\overline{D}) T} T B_{c,2} \overline{V}_\mathrm{in}, \label{eq:sys_dt_buck_mat_a} \\
        &B_{\nu_1} = (e^{A_c T}-I) A_c^{-1} B_{c,1}, \\
        &B_{\nu_2} = e^{A_c T} (I - e^{-A_c \overline{D} T}) A_c^{-1} B_{c,2}, \\
        &b = e^{A_c(1-\overline{D})T} \left[\big(e^{A_c\overline{D}T}-I\big)A_c^{-1}-T\overline{D}\,I\right] B_{c,2} \overline{V}_{\mathrm{in}}, \\
        &C = C_c, \quad D_{\nu_1} = D_{c,1}.
    \end{align}
\end{subequations}

\begin{remark}
    Note that, since the linearization is performed around the nominal values $\overline{D}$ and $\overline{V}_\mathrm{in}$, the equilibrium point $(\overline{x}, \overline{D}, \overline{V}_o)$ of the nonlinear system~\eqref{eq:sys_dt_buck_nl}, is also an equilibrium of the linearized system in the undisturbed case. Thus, Assumption~\ref{ass:mpc_ref_equil} is met.
\end{remark}

\subsection{Output Current Disturbance and State Estimation} \label{sec:estim_design}

According to the setup in Sec.~\ref{sec:probl_formul}, both external disturbances and states must be either measured or estimated to apply EMPC design. This section describes a simple, low-complex design for devices that estimate $\nu(t)$ and $x(t)$.

\subsubsection{Disturbances Estimation}

The vector of disturbances is $\nu(t) = [i_{o}(t), v_{\mathrm{in}}(t)]^\top$. Since $V_\mathrm{in}(t)$ is available for measurement, $\nu_2(t) = v_{\mathrm{in}}(t)$ is trivially obtained as $v_{\mathrm{in}}(t) = V_{\mathrm{in}}(t) - \overline{V}_\mathrm{in}$ and does not need to be estimated. On the other hand, the output current drawn from the load, $\nu_1(t) = i_{o}(t)$, is not directly available for measurement; therefore, it must be estimated. In the remainder of this section, we describe how to design a device producing an estimate $\hat{i}_o(t)$ of $i_o(t)$. Specifically, we resort to the linear estimator proposed in~\cite{gab25_estimator}, based on algebraic design. Given $v_o(t)$ and $i_L(t)$, the problem of estimating $i_o$ is a linear algebraic problem. Indeed, it is obtained in~\cite{gab25_estimator} that $i_o(s) = -E_1(s) v_o(s)  + i_L(s)$ where
\begin{align}\label{eq:defG1}
    E_1(s) \doteq \frac{P_{21}(s)}{P_{11}(s)} = \frac{C_o(R_L+R_{C_o})s+1}{R_L(1+C_oR_{C_o} s)}.
\end{align}
We refer the reader to~\cite{gab25_estimator} for the complete proof and a detailed discussion. 
Consequently, given the measured $v_o(t)$ and the measurement of the inductor current $i_{L,\mathrm{sc}}(t) = g_{i_L} i_L(t)$ (where $g_{i_L} \in \mathbb{R}_{>0}$ is the corresponding sensor gain), the output current estimate $\hat{i}_o(t)$ is obtained by:
\begin{align} \label{eq:estim_io_expr} 
    \hat{i}_o(s) = \mathcal{E}(s) \begin{bmatrix}v_o(s) \\ i_{L,\mathrm{sc}}(s)\end{bmatrix}, \quad \mathcal{E}(s) \doteq \begin{bmatrix} -E_1(s) & {1}/{g_{i_L}} \end{bmatrix}.
\end{align}
\begin{remark}
    The design of the estimator is entirely conducted in continuous time (CT) and relies on signals involved in the linear subsystem only. Consequently, the problem is solved exactly by the CT LTI filter $\mathcal{E}(s)$.
\end{remark}

In ideal conditions, $\hat{i}_o(t) = i_o(t)$ for all $t \in \mathbb{R}_{\geq 0}$, but the presence of noise and modeling uncertainties in the plant description may hinder the quality of the estimate.
Regarding noise, an appropriately designed printed circuit board (PCB) can minimize parasitic effects that perturb the measured signals before they are processed by $\mathcal{E}$.
Regarding modeling uncertainties, we note that only $R_L$, $C_o$, and $R_{C_o}$ are involved in the design; thus, uncertainties in any other components are irrelevant.
 We study the impact of the uncertainties on $R_L$, $C$, and $R_{C_o}$ through the following sensitivity analysis:
\begin{subequations}
\begin{align}
    S^{E_1}_{C}(s) \doteq \frac{\partial E_1}{\partial C} &= \frac{1}{(C R_{C_o})^2} \frac{s}{(s + \frac{1}{C R_{C_o}})^2}, \\
    S^{E_1}_{R_{C_o}}(s) \doteq \frac{\partial E_1}{\partial R_{C_o}} &= -\frac{1}{R_{C_o}^2} \frac{s^2}{(s + \frac{1}{C R_{C_o}})^2}, \\
    S^{E_1}_{R_L}(s) \doteq \frac{\partial E_1}{\partial R_L} &= -\frac{1}{R_L^2}.
\end{align}
\end{subequations}
As $s \rightarrow 0$, both $\vert S^{E_1}_{C}\vert \rightarrow 0$ and $\vert S^{E_1}_{R_{C_o}}\vert \rightarrow 0$, indicating that uncertainty on $C$ and $R_{C_o}$ poorly influence the estimate $\hat i_o$ at low frequencies. Instead, $\vert S^{E_1}_{R_L}\vert $ is non-zero, indicating that the uncertainty of $R_L$ leads to some steady-state estimation error $\vert \hat i_o(t) - i_o(t) \vert$.

Let $\hat{C_o}$, $\hat{R}_{C_o}$, and $\hat{R}_L$ denote the nominal values of these components and $\hat{E}_1$ the transfer function $E_1$ in Eq.~\eqref{eq:defG1} when evaluated using $\hat{C_o}$, $\hat{R}_{C_o}$, and $\hat{R}_L$. To counteract the effect of the uncertainty on $R_L$, we propose a robust design of $\hat{E}_1$ to minimize the expected value of the steady-state estimation error $|\hat{i}_o(t) - i_o(t)|$. At steady state, since $v_o$ is constant, $v_o(s) = \overline{V}_o/s$ and by the final value theorem, we obtain:
\begin{align}
    & \hspace{-10pt} \lim_{t \rightarrow \infty} \left|\hat{i}_o(t) - i_o(t)\right| = \lim_{s \rightarrow 0} \left|s \left(\hat{i}_o(s) - i_o(s)\right)\right| \nonumber \\
    & \hspace{-10pt} \quad = \lim_{s \rightarrow 0} \left| s \left(\hat{E}_1(s) - E_1(s)\right) \frac{\overline{V}_o}{s} \right| = \overline{V}_o \left| \frac{1}{\hat{R}_L} - \frac{1}{R_L} \right|.
\end{align}
First, we notice that such an error only depends on the uncertainty on $R_L$, and not on that on $C_o, R_{C_o}$, which only affects the error transient: for a frequency-dependent sensitivity analysis of the error, see the extended version of this manuscript\hyperlink{ref:ext_ver}{\footnotemark[1]}. Formally, we establish the following result:
\begin{proposition}[Optimal estimator $\hat{R}_L^*$]
    Let $R_L$ be a uniformly distributed random variable in $[R_{L,\mathrm{min}}, R_{L,\mathrm{max}}]$, i.e., $R_L \sim U([R_{L,\mathrm{min}}, R_{L,\mathrm{max}}])$.
    Then, the solution to
    \begin{align} \label{eq:opt_RLhat_probl}
        \hat{R}_L^* = \argmin_{\hat{R}_L \in \mathbb{R}} \mathbb{E}\left[ \left\vert \frac{1}{\hat{R}_L} - \frac{1}{R_L} \right\vert \right],
    \end{align}
    is given by the mean resistance $\hat{R}_L^* = \frac{1}{2}{\left(R_{L,\mathrm{min}} + R_{L,\mathrm{max}}\right)}.$
\end{proposition}
\begin{proof}
   Consider the change of coordinates $a = 1/\hat{R}_L$ and $B = 1/R_L$. The probability density function of $B$ is given by the change of variables formula~\cite[Sec.~2.3]{bookcasella}:
    \begin{align}
        \hspace{-8pt} f_B(b) \! = \! \begin{cases}
            \frac{1}{R_{L,\mathrm{max}}-R_{L,\mathrm{min}}} b^{-2} & \textrm{if} \;\; \frac{1}{R_{L,\mathrm{min}}} \leq b \leq \frac{1}{R_{L,\mathrm{max}}}, \\
            0 & \textrm{otherwise}.
        \end{cases}
    \end{align}
    After the change of variables, the problem takes the form
    \begin{align} \label{eq:opt_RLhat_probl_2}
        \hat{a}^* = \argmin_{a \in \mathbb{R}} \mathbb{E}\left[ \left\vert a-B \right\vert \right].
    \end{align}
    It is a standard result that the optimum of such a problem corresponds to the median of the distribution of $B$~\cite[Chapter~1]{casellabook2}, i.e., $a^*$ satisfies
    $
        \int_{-\infty}^{a^*} f_B(b) db = \frac{1}{2}.
    $
    Computing the integral, we get
    \begin{align}
        \frac{1}{2} = \frac{1}{R_{L,\mathrm{max}}-R_{L,\mathrm{min}}} \left[ -\frac{1}{b} \right]_{R_{L,\mathrm{max}}^{-1}}^{a^*} = \frac{R_{L,\mathrm{max}}-a^*}{R_{L,\mathrm{max}}-R_{L,\mathrm{min}}}.
    \end{align}
    Solving for $a^*$ and taking its inverse yields the result.
\end{proof}

The estimator $\mathcal{E}(s)$, defined in Eq.~\eqref{eq:estim_io_expr}, allows for a cheap circuit implementation as it is composed of a first-order filter $E_1(s)$ and a gain only. 

\begin{figure}[t]
    \centering
    \includegraphics[width=\linewidth]{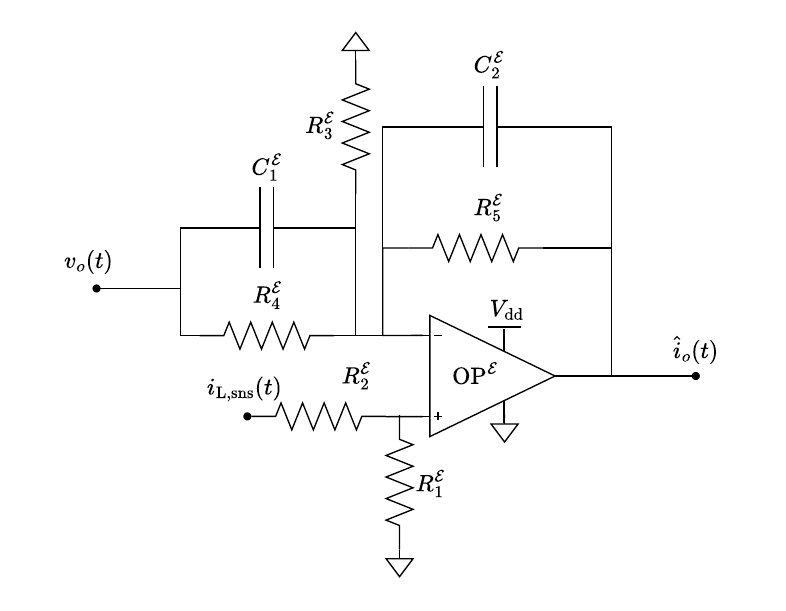}
    \caption{Estimator circuit.}
    \label{fig:implem_Estim}
\end{figure}

We start from the definition of $E_1(s)$ in Eq.~\eqref{eq:defG1}, which can be equivalently rewritten as
\begin{align}
    E_1(s) &= 
    \frac{1}{R_L} \frac{1 + {s}{C_o(R_L + R_{C_o})}}{1 + {s}{C_o R_{C_o}}}\\ &= K_\mathcal{E} \frac{1 + {s}/{z_\mathcal{E}}}{1 + {s}/{p_\mathcal{E}}},
\end{align}
where 
\begin{align}
    K_\mathcal{E} = \frac{1}{R_L}, \quad z_\mathcal{E} = \frac{1}{C_o(R_L + R_{C_o})}, \quad p_\mathcal{E} = \frac{1}{C_o R_{C_o}}.
\end{align}
Since $R_{C_o}+R_L \gg R_{C_o}$, we have that $z_\mathcal{E}<p_\mathcal{E}$ and $E_1(s)$ is a high-pass filter. Consequently, we propose an analog implementation based on the circuital topology in Fig.~\ref{fig:implem_Estim} (highlighted in green in Fig.~\ref{fig:sim_circ}), which differs from the LEC implementation considered in~\cite{gab25_estimator} due to the absence of feedforward compensation. To avoid that the output of the OP-AMP $\mathrm{OP}^\mathcal{E}$ exceeds the saturation limits, we scale the gain of $\mathcal{E}$ by a factor $g_{i_o} \in (0,1)$, obtaining a scaled version of the current estimate $\hat{i}_{o,\mathrm{sc}} = g_{i_o} \hat{i}_o$.

The transfer functions of the circuit in Fig.~\ref{fig:implem_Estim}, from inputs $v_o$ and $i_{L,\mathrm{sns}}$ to output $\hat{i}_o$ are 
\begin{subequations} \label{eq:circ_eq_est}
    \begin{align}
        \frac{\hat{i}_{o,\mathrm{sc}}(s)}{v_o(s)} &= -\frac{R_5^\mathcal{E}}{R_4^\mathcal{E}}\frac{1 + sR_4^\mathcal{E}C_1^\mathcal{E}}{1 + sR_5^\mathcal{E}C_2^\mathcal{E}}, \label{eq:circ_eq_est_a} \\
        \frac{\hat{i}_{o,\mathrm{sc}}(s)}{i_{L,\mathrm{sc}}(s)} &= \frac{R_1^\mathcal{E}}{R_1^\mathcal{E}+R_2^\mathcal{E}} \frac{1 + R_5^\mathcal{E}(R_3^\mathcal{E} \parallel R_4^\mathcal{E})^{-1} + sR_5(C_1+C_2)}{1 + s R_5 C_2}. \label{eq:circ_eq_est_b}
    \end{align}
\end{subequations}
Matching Eq.~\eqref{eq:circ_eq_est} with the desired estimator transfer function in Eq.~\eqref{eq:estim_io_expr} requires imposing the conditions 
\begin{align}
    g_{i_o} K_\mathcal{E}=R_5^\mathcal{E}/R_4^\mathcal{E}, p_\mathcal{E}=(R_5^\mathcal{E}C_2^\mathcal{E})^{-1}, z_\mathcal{E}=(R_4^\mathcal{E}C_1^\mathcal{E})^{-1}
\end{align}
to implement the $E_1(s)$ transfer function. Then, we select 
\begin{align}
    R_4^\mathcal{E} = R_2^\mathcal{E}, \quad R_5^\mathcal{E} = g_{i_L}^{-1} g_{i_o} R_1^\mathcal{E}
\end{align}
and $R_3^\mathcal{E}$ such that
\begin{equation}
    C_1^\mathcal{E}(R_3^\mathcal{E}~\parallel~R_4^\mathcal{E}) = C_2^\mathcal{E} R_5^\mathcal{E}
\end{equation} 
to ensure that $\hat{i}_o(s) / i_{L,\mathrm{sns}}(s) = g_{i_{L}}^{-1} g_{i_o}$. 

\subsubsection{State Estimation}
Since $i_L(t)$ can easily be sensed (see, e.g.,~\cite{gab25_estimator}), but the capacitor voltage $v_C(t)$ is not accessible due to the presence of the capacitor equivalent series resistance (ESR), the estimation of $v_C(t)$ is needed. This section presents a simple algebraic estimator design for $v_C(t)$. 
Starting from Kirchhoff's current law on the output node of $\mathcal{S}_2$, we have
\begin{align}
    i_C = i_L - i_o - v_o/R_L = (v_o - v_C)/R_{C_o}.
\end{align}
Then, given the measurements of $i_L$ and $v_o$ and the estimate $\hat{i}_o$ of $i_o$, we define the estimator
\begin{align} \label{eq:estim_vc}
    \hat v_C = {R_{C_o}} (\hat{i}_o-i_L) +(R_{C_o}/R_L+1)\,v_o.
\end{align}

With this procedure, $\hat v_C(t)$ is obtained using only weighted sums. Consequently, compared to the conventional observer-based strategy, the proposed approach streamlines the design and enables a low-cost analog implementation.

\begin{remark}
    In many practical applications, $R_{C_o} \ll R_L$. Under this assumption, Eq.~\eqref{eq:estim_vc} can be simplified to the straightforward relation $\hat{v}_C = v_o$, meaning that the controlled output voltage can be directly used as a reliable estimate of the state ${v}_C(t)$. This further simplifies the circuit design and reduces both the economic costs and the required board area.
\end{remark}

\subsection{EMPC Design} \label{sec:buck_empc_design}

We construct the EMPC policy according to Sec.~\ref{sec:probl_formul} and~\ref{sec:explicit_mpc}, yielding the following expression:
\begin{align} \label{eq:buck_empc_law_true}
    u_k = \; & d_k = \pi(p_k) = \pi([i_{L,k}, \; \hat{v}_{C,k}, \; \hat{i}_{o,k}, \; v_{\mathrm{in},k}]^\top) \nonumber \\
    = \; & \pi\left(\left[\tfrac{1}{g_{i_L}} i_{L,\mathrm{sc},k}, \; \hat{v}_{C,k}, \; \tfrac{1}{g_{i_o}} \hat{i}_{o,\mathrm{sc},k}, \; v_{\mathrm{in},k}\right]^\top\right).
\end{align}
Recalling Eq.~\eqref{eq:mpc_op_d}, we only impose upper-lower bound constraints on the input, i.e., the duty cycle $d$, since, by definition, it can only range between $0$ ($0\%$) and $1$ ($100\%$). This yields the constraint matrices $H_u = [1, \; -1]^\top$ and $h_u = [1, \; 0]^\top$.

We conduct MPC controller design according to Sec.~\ref{sec:probl_formul} and~\ref{sec:explicit_mpc}. Specifically, within the MPC formulation in Eq.~\eqref{eq:mpc_op}, we utilize the linearized Buck model described by matrices according to Eq.~\eqref{eq:sys_dt_buck_mat} and include input saturation limits $u_\mathrm{lb} \leq \hat{u}_i \leq u_\mathrm{ub}$ as polytopic constraints \eqref{eq:mpc_op_d}, with limits $u_\mathrm{lb}=$\SI{0}{\volt} and $u_\mathrm{ub}=$\SI{1}{\volt} representing the PWM duty-cycle limits. The selected prediction horizon is $N_p=5$.
Regarding the weights $Q, R, R_\Delta$, we employ a trial-and-error procedure to optimize the qualitative observation of load and line disturbance responses. Finally, to reduce the number of regions, we apply the strategies introduced in Sec.~\ref{sec:explicit_mpc}; specifically, in the selection of the move blocking strategy map in Eq.~\eqref{eq:mpc_move_block_matrix}, we select control horizon $N_c=2$.

\subsection{Stability Analysis} \label{sec:stability}

In this section, we establish the existence and local stability of an equilibrium state $\overline{x} \in \mathbb{R}^2$ for the closed-loop system defined by the nonlinear DT Buck converter model~\eqref{eq:sys_dt_buck_nl} and the EMPC policy~\eqref{eq:buck_empc_law_true}. 

The following result establishes local stability robustly to the presence of bounded, constant load and line disturbances.  

\begin{theorem}[Local Stability]
    Let us consider constant disturbances $i_{o,k} = i_o$ and $v_{\mathrm{in},k} = {v}_{\mathrm{in}}$ for all $k \geq 0$.
    Then, the closed-loop system~\eqref{eq:sys_dt_buck_nl},~\eqref{eq:buck_empc_law_true} admits an equilibrium state $\overline{x}$.
    
    Further, let $Z = e^{A_c T (\overline{D} - \overline{u})} = A^{\overline{D} - \overline{u}}$, with $\overline{D}$ defined as in Eq.~\eqref{eq:equil_explicit} for the undisturbed case, and $\overline{u} = \pi(\overline{p})$, with $\overline{p} = [\overline{x},i_o,v_\mathrm{in}]^\top$, according to Eq.~\eqref{eq:buck_empc_law_true}. Assume there exists a matrix $P \in \mathbb{R}^{2\times 2}$ such that $\norm{A+B K_r}_P \leq 1 - \epsilon_s$ with $0<\epsilon_s<1$ and $K_r \in \mathbb{R}^2$ is the vector of the first two elements of the gain $K_i$ of region $\mathcal{R}_i$ such that $[\bar x, 0,0]^\top \in \mathcal{R}_i$. If the disturbances $i_o$ and $v_\mathrm{in}$ are such that
    \begin{equation} \label{eq:cond_th_stab}
        \norm{Z-I}_P + \epsilon_V \norm{Z}_P \leq \frac{\epsilon_s}{\norm{B K_r}_P},
    \end{equation}
    where $\epsilon_V = v_\mathrm{in}/\overline{V}_\mathrm{in}$, then the equilibrium $\overline{x}$ is locally stable.
\end{theorem}

\begin{proof}
    First, we prove the existence of the equilibrium state $\overline{x}$ satisfying
    \begin{equation}\label{eq:cond_nl_th_eq}
        (A-I)\overline{x} + B_{\nu,2} i_o + g(\pi(\overline{p}), V_\mathrm{in}) = 0.
    \end{equation}
    As established in~\cite[Corollary~2.2.2]{dinca2009brouwer}, if $A-I$ is invertible and $g(\pi(\overline{p}), V_\mathrm{in})$ is continuous and bounded, then Eq.~\eqref{eq:cond_nl_th_eq} admits at least one solution. This holds since $0 \leq u(x) \leq 1$ implies that $g(u(\overline{x}), V_\mathrm{in})$ remains bounded.
    
    We now prove the local stability of $\overline{x}$. First, notice that in a neighbourhood of $\overline{p}$ the EMPC policy reduces to $\pi(x) = K_i [x,i_o,v_\mathrm{in}]^\top + l_i = K_r x + l_r$ for $[x,i_o,v_\mathrm{in}]^\top \in \mathcal{R}_i$, where $\mathcal{R}_i$ is the EMPC region containing $[x,0,0]^\top$, $K_r \in \mathbb{R}^2$ and $l_r = l_i + K_{i,3}i_o + K_{i,4}v_\mathrm{in} \in \mathbb{R}$. Then, the closed-loop system is equivalently written as
    \begin{align} \label{eq:closedloop_proof_stab}  
        x_{k+1} = F(x_k) = A x + B_{\nu,2} i_o + g(K_r x_k + l_r, V_\mathrm{in}).
    \end{align}
    
    We show that Eq.~\eqref{eq:closedloop_proof_stab} locally defines a contraction in the $P$-norm, which is equivalent to the following condition (see, e.g.,~\cite{bullo2022contraction} for more details):
    \begin{align}
        \norm{J(\overline{x})}_P < 1, \quad J(\overline{x}) = \frac{\partial F}{\partial x}(\overline{x}).
    \end{align}
    For system~\eqref{eq:closedloop_proof_stab}, we obtain
    \begin{align}
        J(\overline{x}) = A + e^{A_c ( 1 - K_r \overline{x} - l_r) T} T B_{c,2} V_\mathrm{in} K_r.
    \end{align}
    Recalling that $B = e^{A_c ( 1 - \overline{D}) T} T B_{c,2} \overline{V}_\mathrm{in}$ from Eq.~\eqref{eq:sys_dt_buck_mat_a}, we can rewrite
    \begin{align}
        J(\overline{x}) = A + {\Delta} B K_r, \quad \Delta = e^{A_c (\overline{D} - \overline{u}) T} \frac{V_\mathrm{in}}{\overline{V}_\mathrm{in}}.
    \end{align}
    
    A sufficient condition for $\norm{J(\overline{x})}_P < 1$ is then
    \begin{equation}\begin{aligned}
        \norm{A+\Delta BK_r}_P &\leq \norm{A+BK_r}_P+\\
        &+\norm{\Delta-I}_P\norm{BK_r}_P< 1.
    \end{aligned}\end{equation}
    Since, by assumption, $\norm{A+BK_r}_P \leq 1 - \epsilon_s$, we obtain
    \begin{align} \label{eq:th_local_cond_Delta_1}
        \norm{\Delta-I}_P \leq \frac{\epsilon_s}{\norm{BK_r}_P}.
    \end{align}
    Then, recalling the expressions of $\Delta$ and $Z$, and requiring that $v_\mathrm{in} = \overline{V}_\mathrm{in}\epsilon_V$, yields
    \begin{equation}\begin{aligned}\label{eq:th_stab_final}
        \norm{\Delta-I}_P &= \norm{Z (1+\epsilon_V)-I}_P \leq\\
        &\leq \norm{Z-I}_P + \epsilon_V \norm{Z}_P.
    \end{aligned}\end{equation}
    Finally, replacing Eq.~\eqref{eq:th_stab_final} into Eq.~\eqref{eq:th_local_cond_Delta_1}, we obtain the condition in Eq.~\eqref{eq:cond_th_stab}.
\end{proof}

The local feedback matrix $K_r$ coincides with the standard linear quadratic regulator gain when a sufficiently long prediction horizon and no move blocking are considered. Under these ideal conditions, $A+B K_r$ is guaranteed to be Schur stable. Thus, the existence of a matrix $P$ and a scalar $\epsilon_s>0$ such that $\norm{A+B K_r}_P = 1 - \epsilon_s$ is guaranteed. In our setting, the adopted move blocking strategy induces only mild perturbations compared to the unblocked formulation, so the resulting closed-loop matrix is expected to remain stabilizing. In practice, the Schur stability property of $A+B K_r$ can still be verified a posteriori.

\begin{remark}
    The Shur stability property of $A+B K_r$ serves only as a preliminary condition, and our local stability certificate is a stronger result. Beyond asymptotic stability, our analysis quantifies the admissible disturbance set for which the nonlinear closed-loop system remains stable, by computing $\epsilon_s$, $\epsilon_V$, and $Z$. Specifically, computing $\epsilon_s$ and $\epsilon_V$ is trivial, while $\norm{Z}_P$ and $\norm{Z-I}_P$ are monotone functions of $\overline{D} - \overline{u}$: we compute $\overline{D}$ from Eq.~\eqref{eq:equil_explicit} and $\overline{u}$ from Eq.~\eqref{eq:cond_nl_th_eq} using the worst-case value of the disturbances. For small disturbance amplitudes, we have $Z \approx I$ and $\epsilon_V \approx 0$, making the condition~\eqref{eq:cond_th_stab} trivially satisfied.
\end{remark}

\section{Simulations and Results} \label{sec:results}

In this section, we validate our circuital EMPC approach, applied to Buck converters, through extensive simulations.
Specifically, we consider two simulation scenarios:
\begin{enumerate}[label={\arabic*)}, leftmargin=*]
    \item high-level Monte Carlo simulations performed in \textsc{Matlab}\textsuperscript{®} and Simulink\textsuperscript{®} (ver. 2023b) to assess robust control performance in the presence of parametric uncertainty;
    
    \item high-fidelity circuit-level simulations performed in LTSpice\textsuperscript{®}~\cite{ltspice} to assess the control performance under realistic operating conditions.
\end{enumerate}



%


The relevant data, shared by all simulations, are as follows:
\begin{itemize}[label={$\bullet$}, leftmargin=*]
    \item General data: $T = 2\;\unit{\micro s}$, $f_\mathrm{sw} = \frac{1}{T} = 500\,\unit{kHz}$.
    
    \item Buck converter: $\overline{V}_\mathrm{in} = 50\;\unit{V}$ (nominal), $V_{\mathrm{in}} \in [25, \; 60]\;\unit{V}$ (range); $\overline{V}_o = 5\;\unit{V}$; $I_{o,\mathrm{max}} = 15\;\unit{A}$; $R_L = 3.681\;\unit{\ohm}$ (nominal), $R_L \in [0.333, \; 7.029]\;\unit{\ohm}$ (range); $C_o = 250\;\unit{\micro F}$ (nominal), $C_o \in 250\;\unit{\micro F} \pm 10\% = [225, \; 275]\;\unit{\micro F}$ (uncertainty); $L = 8.2\;\unit{\micro H}$ (nominal), $L \in 8.2\;\unit{\micro H} \pm 20\% = [6.56, \; 9.84]\;\unit{\micro H}$ (uncertainty).
    \begin{itemize}[label={$\circ$}, leftmargin=*]
        \item Ceramic capacitor: $R_{C_o} = 5\;\unit{m\ohm}$ (nominal), $R_{C_o} \in 5\;\unit{m\ohm} \pm 50\% = [2.5, \; 7.5]\;\unit{m\ohm}$ (uncertainty).
        
        \item Electrolytic capacitor: $R_{C_o} = 50\;\unit{m\ohm}$ (nominal), $R_{C_o} \in 50\;\unit{m\ohm} \pm 50\% = [25, \; 75]\;\unit{m\ohm}$ (uncertainty).
    \end{itemize}
    
    \item Estimators and sensors: $g_{i_L} = 0.2\;\unit{V.A^{-1}}$, $g_{i_o} = 0.1$.
    
    \item MPC: $N_p = \num{5}$, $N_c = \num{2}$, $Q = 10^2$, $R = 10^{-2}$, $R_\Delta = 1$.
    
    \item EMPC: $\mathcal{P} = \{p = (i_L, \hat{v}_C, \hat{i}_o, v_\mathrm{in}) \in \mathbb{R}^4: i_L \in [0, \; 80]\;\unit{A}, \; \hat{v}_C \in [0, \; 20]\;\unit{V}, \; \hat{i}_o \in [-5, \; 20]\;\unit{A}, \; v_\mathrm{in} = [- 35\;\unit{V}, \; 35\;\unit{V}]\}$.
\end{itemize}

Concerning the capacitor ESR $R_{C_o}$, we highlight that a very large uncertainty is considered.

\subsection{EMPC Design and Complexity Reduction} \label{sec:buck_empc_design_results}

\begin{figure*}[t!]
    \centering
    \includegraphics[width=\linewidth]{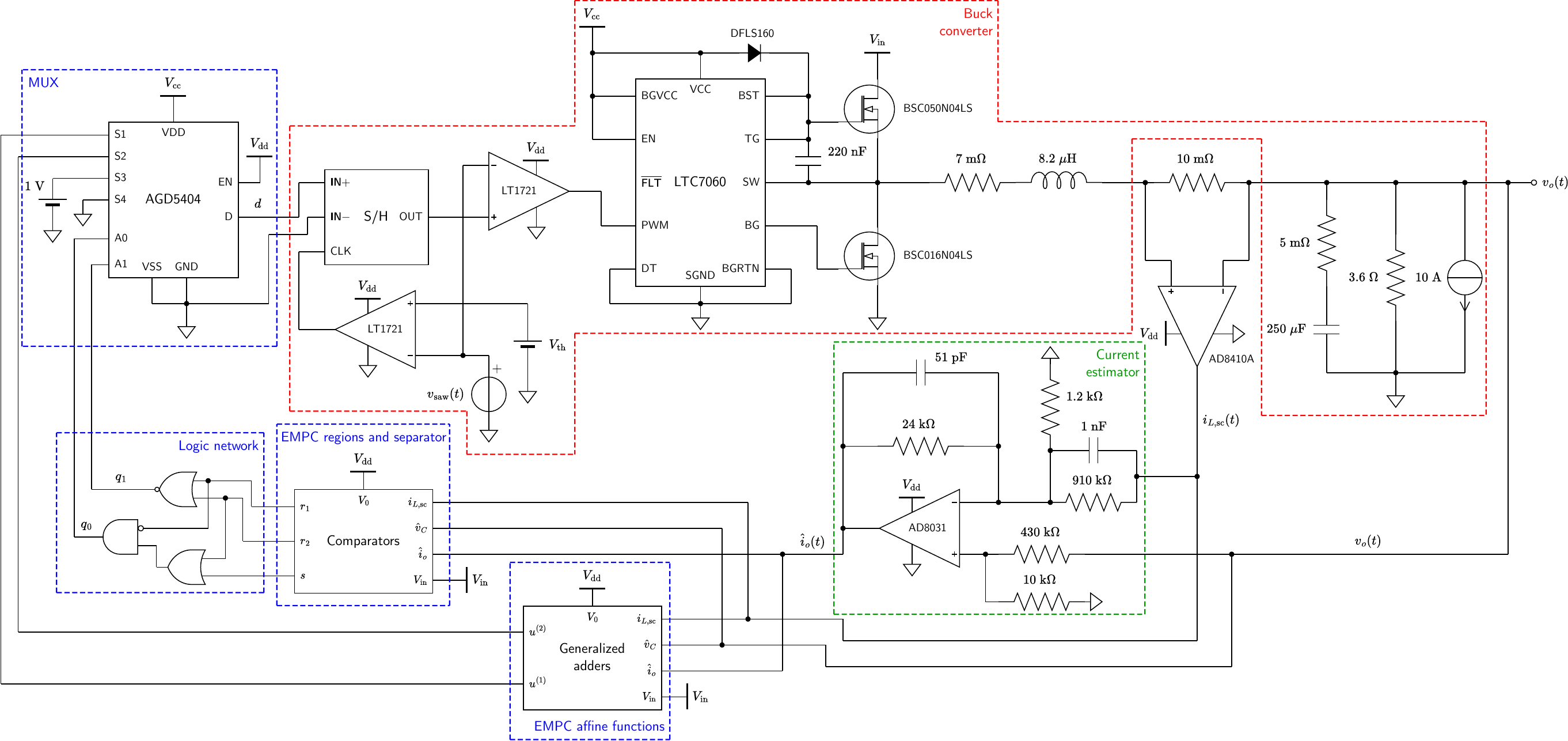}
    \caption{EMPC analog circuit controlling the Buck converter: complete schematic.}
    \label{fig:sim_circ}
    \vspace{-0.5\baselineskip}
\end{figure*}

\begin{figure}[t!]
    \begin{minipage}[t]{0.475\linewidth}
        \centering
        \caption*{(a)}
        \includegraphics[width=\linewidth]{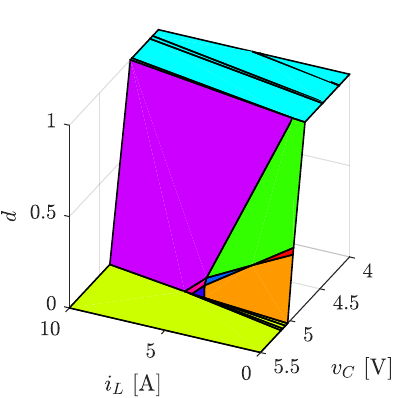}
    \end{minipage}
    \hfill
    \begin{minipage}[t]{0.475\linewidth}
        \centering
        \caption*{(b)}
        \includegraphics[width=\linewidth]{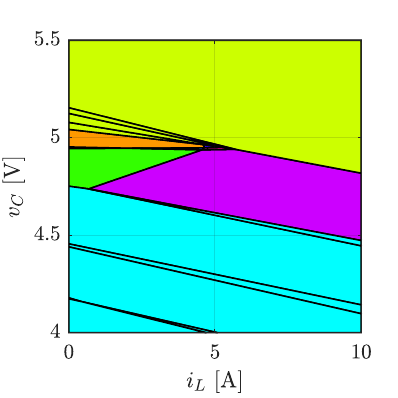}
    \end{minipage}


    \begin{minipage}[t]{0.475\linewidth}
        \centering
        \caption*{(c)}
        \includegraphics[width=\linewidth]{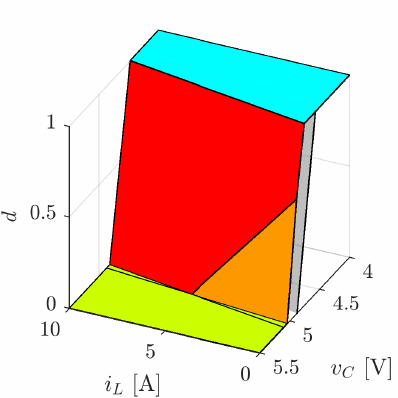}
    \end{minipage}
    \hfill
    \begin{minipage}[t]{0.475\linewidth}
        \centering
        \caption*{(d)}
        \includegraphics[width=\linewidth]{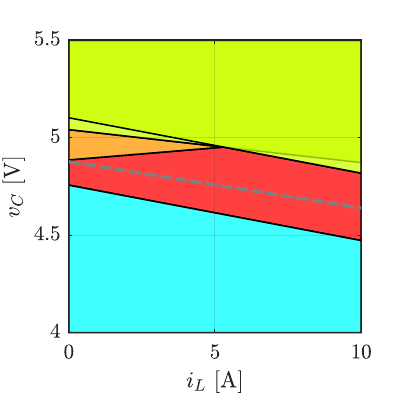}
    \end{minipage}
    \caption{EMPC policy for the Buck converter, before (a, b) and after (c, d) complexity reduction. In (c) and (d), it is also reported the affine separation function $\sigma$~(\tikzreffill[fill=black!50!white, draw=black, line width=0.5pt, fill opacity=0.5]) and its zero-level set~(\tikzref[color=black!50!white, line width=1.5pt, dash pattern=on 4pt off 1pt]). \tikzreffill[fill=yellow!60, draw=black, line width=0.5pt, fill opacity=0.5] and \tikzreffill[fill=cyan!40, draw=black, line width=0.5pt, fill opacity=0.5] denote regions saturated at $0$ and $1$, respectively.}
    \vspace{-\baselineskip}
    \label{fig:buck_empc_law}
\end{figure}

We derive the EMPC policy \eqref{eq:buck_empc_law_true} following Sec.~\ref{sec:explicit_mpc} and~\ref{sec:buck_empc_design} by using nominal component values. Let us consider the case with the ceramic capacitor (i.e., $R_{C_o} = 5\;\unit{m\ohm}$, analogous considerations apply to the electrolytic case). The resulting EMPC policy $\pi$ is composed of $R = 19$ regions ($7$ unsaturated, $12$ saturated), with $9$ unique affine functions ($7$ unsaturated, $1$ saturated to $u_\mathrm{lb}$, $1$ saturated to $u_\mathrm{ub}$).
We reduce the complexity of $\pi$ by applying the four techniques reported in Sec.~\ref{sec:explicit_mpc} in the following order:
\begin{enumerate}[leftmargin=*]
    \item move blocking strategy simplifies the EMPC policy to $R=7$ regions ($2$ unsaturated, $5$ saturated) with $4$ unique affine functions ($2$ unsaturated, $2$ saturated).
    
    \item the non-disjoint optimal merging of regions leads to $R = 5$ regions ($2$ unsaturated, $3$ saturated) with $4$ unique affine functions ($2$ unsaturated, $2$ saturated). 
    
    \item the hyperplane separation of saturated regions further reduces to only $2$ unsaturated regions, with $2$ unique unsaturated affine functions. 

    \item To remove trivial inequalities, we observe that the $2$ unsaturated regions have $1$ common facet and are delimited by $6$ hyperplanes, $2$ of which are given by the set $\mathcal{P}$. As a consequence, the unsaturated regions of the simplified EMPC policy can be defined by $4$ non-trivial inequalities only, of which one is shared.
\end{enumerate}
Overall, the number of regions to be checked (i.e., unsaturated) drops from $19$ to $2$, corresponding to a $89\%$ reduction. Fig.~\ref{fig:buck_empc_law} shows the EMPC policy before and after complexity reduction. The control law is projected onto a reduced two-dimensional state space with $\{(i_L, \hat{v}_C) \in \mathbb{R}^2: i_L \in [0, \; 10]\;\unit{A}, \; \hat{v}_C \in [4, \; 5.5]\;\unit{V}\}$, while holding $\hat{i}_o = 0\;\unit{A}$ and $V_\mathrm{in} = \overline{V}_\mathrm{in} = 50\;\unit{V}$.

\subsubsection{Comparison With the Literature}

Compared to the recent work~\cite{wisniewski_explicit_2021} on EMPC control with region elimination for Buck converters, we observe that our approach yields as few as $4$ regions while considering a very fast switching frequency $f_\mathrm{sw}=\SI{500}{\kilo\hertz}$; conversely,~\cite{wisniewski_explicit_2021} reports an increase from $3$ to $7$ regions when increasing $f_\mathrm{sw}$ from \SI{4}{\kilo\hertz} to \SI{10}{\kilo\hertz}, potentially leading to an overly complex control law at the desired frequency of \SI{500}{\kilo\hertz}.

\subsection{Circuital Implementation} \label{sec:empc_circ_implem}

The analog circuit implementing the complexity-reduced EMPC policy is realized according to Sec.~\ref{sec:circuit} and the estimator $\mathcal{E}$ is designed according to Sec.~\ref{sec:estim_design}. For passive circuit components, we use values from the E-series standard. The overall circuit schematic is reported in Fig.~\ref{fig:sim_circ}.

The power stage of the Buck converter is implemented through the half-bridge gate driver \texttt{LTC7060}, which drives the high-side and low-side power MOSFETs of the half-bridge, which are the \texttt{BSC050N04LS} and \texttt{BSC016N04LS}, respectively, depending on the state of the logic signal provided on the PWM input pin.
The Dickson's charge pump, including $C_\mathrm{pump}$-$D_\mathrm{pump}$, serves to drive the gate of the n-channel MOSFET \texttt{BSC050N04LS}, and it takes as input the voltage $V_\mathrm{cc} = 12\;\unit{V}$.
Both the load current estimator (Fig.~\ref{fig:sim_circ}, green box) and the generalized adders, which implement the EMPC affine functions (Fig.~\ref{fig:genadd}), utilize the \texttt{AD8031} OP-AMP: a rail-to-rail, single-supply OP-AMP characterized by $80\;\unit{\mega\hertz}$ gain-bandwidth product. 
The comparator used to implement the EMPC regions and the affine separator (Fig.~\ref{fig:cmp}) is the single-supply \texttt{LT1721}.
The inductor current sensing is performed through a $10\;\unit{\milli\ohm}$ shunt resistance $R_\mathrm{sense}$, together with the current-sense amplifier \texttt{AD8410A}. The latter offers a $2.2\;\unit{\mega\hertz}$ bandwidth (four times $f_\mathrm{sw}$) and is configured so that its voltage gain is $A_\mathrm{sense}=20$. Therefore, the overall gain is $g_{i_L} = R_\mathrm{sense}A_\mathrm{sense} = 0.2\;\unit{V.A^{-1}}$. The OP-AMPs, the current-sense amplifier, and the comparators are supplied with a voltage $V_\mathrm{dd} = 5\;\unit{V}$. Also, within the generalized adders and comparators, we set the constant voltage $V_0 = V_\mathrm{dd} = 5\;\unit{V}$ (refer to Sec.~\ref{sec:circuit}). Finally, the 4-channel analog MUX is the \texttt{ADG5404}, which operates with the $V_\mathrm{cc}$ supply voltage.

\subsubsection{Effect of Complexity Reduction on Circuital Implementation}

After the four-fold complexity reduction carried out in Sec.~\ref{sec:buck_empc_design_results}, we draw conclusions on how many components are required to implement the analog EMPC circuit:
%
\begin{itemize}[label={$\bullet$}, leftmargin=*]
    \item A four-channel MUX, in order to select the control input from each of the $2$ unsaturated affine functions, and the two saturated input values, $d_\mathrm{lb} = 0$ and $d_\mathrm{ub} = 1$.
    
    \item $2$ generalized adders to implement the EMPC affine functions: $1$ for each unique affine function.
    
    \item $5$ comparators: $4$ to implement the non-trivial inequalities describing the EMPC regions ($3$ for the first region and $2$ for the second, with one shared), and $1$ for the affine separator.
\end{itemize}
%

\subsubsection{Logic Gate Network} \label{sec:buck_empc_results_circ_logic}

In order to implement the network of logic gates in Sec.~\ref{sec:circuit_logic_net}, which generates the selection signal $q = (q_0, q_1)$ for the four-channel MUX, we construct a truth table, relating the signals $r_1$, $r_2$, and $s$, according to the simplified EMPC policy and Eqs.~\eqref{eq:law_indicator}--\eqref{eq:empc_policy_final}.
Table~\ref{tab:empc_buck_truth_tab} shows the considered truth table, which correspond to the logic expressions:
\begin{align} \label{eq:buck_empc_logic_net_true}
    q_0 = \overline{r}_1\odot(s \oplus r_2), \quad q_1 = \overline{r_1 \oplus r_2},
\end{align}
which are implemented using only $4$ logic gates: $1$ AND, $1$ OR, $1$ NOR, and $1$ NOT; see Figure~\ref{fig:sim_circ}.
\begin{table}[t!]
    \centering
    \begin{tabular}{ccc|cc}
        \TopRule
        $s$ & $r_1$ & $r_2$ & $q_1$ & $q_0$ \\
        \MidRule
        0   & 0     & 0     & 1     & 0 \\
        0   & 0     & 1     & 0     & 1 \\
        0   & 1     & 0     & 0     & 0 \\
        0   & 1     & 1     & -     & - \\
        1   & 0     & 0     & 1     & 1 \\
        1   & 0     & 1     & 0     & 1 \\
        1   & 1     & 0     & 0     & 0 \\
        1   & 1     & 1     & -     & - \\
        \BottomRule
    \end{tabular}
    \caption{Truth table for the MUX selection signal, implementing the logic functions~\eqref{eq:buck_empc_logic_net_true}. \quotes{-} stands for \quotes{don't care}.}
    \label{tab:empc_buck_truth_tab}
\end{table}

\subsection{Robust Performance Assessment} \label{sec:simulink}
In this section, we assess the robust control performance of the EMPC policy in the presence of uncertainties in the Buck converter model parameters through extensive Monte Carlo simulations. These simulations are performed in the \textsc{Matlab}-Simulink environment due to its higher computational efficiency compared to other, more accurate, circuit simulators.

Within the Buck converter model~\eqref{eq:sys_ct_buck_complete}, we introduce parametric uncertainty on the parameters of its passive components, i.e., $R_L$, $L$, $C_o$, and $R_{C_o}$. In the Monte Carlo simulations, these uncertain parameters are treated as random variables, with a uniform probability distribution within their respective uncertainty intervals.
%
%
For $R_{C_o}$, we consider both the cases of ceramic and electrolytic capacitor.
A total of $500$ random runs are performed.

\begin{figure*}[t!]
    \begin{minipage}[t]{0.475\linewidth}
        \centering
        \caption*{(a) Ceramic capacitor}
        \vspace{-0.5\baselineskip}
        \includegraphics[width=\linewidth]{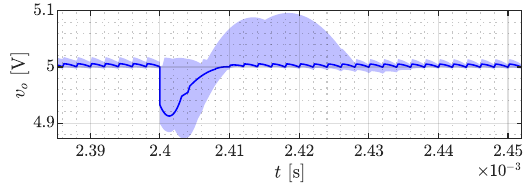}
    \end{minipage}
    \hfill
    \begin{minipage}[t]{0.475\linewidth}
        \centering
        \caption*{(b) Electrolytic capacitor}
        \vspace{-0.5\baselineskip}
        \includegraphics[width=\linewidth]{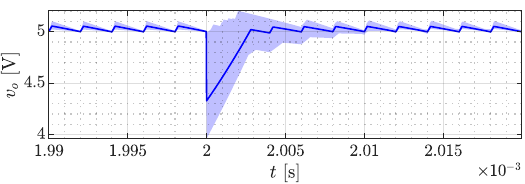}
    \end{minipage}
    \caption{EMPC robust performance: closed-loop system response $v_o(t)$ (nominal~\tikzref[color=blue, line width=0.75pt], uncertainty~\tikzreffill[fill=blue, fill opacity=0.25]) to a step load disturbance $i_o(t)$.}
    \vspace{-0.5\baselineskip}
    \label{fig:buck_empc_sim_mc_load}
\end{figure*}

\begin{figure*}[t!]
    \begin{minipage}[t]{0.475\linewidth}
        \centering
        \caption*{(a) Ceramic capacitor}
        \vspace{-0.5\baselineskip}
        \includegraphics[width=\linewidth]{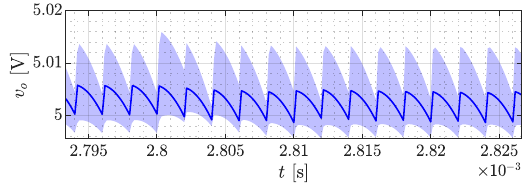}
    \end{minipage}
    \hfill
    \begin{minipage}[t]{0.475\linewidth}
        \centering
        \caption*{(b) Electrolytic capacitor}
        \vspace{-0.5\baselineskip}
        \includegraphics[width=\linewidth]{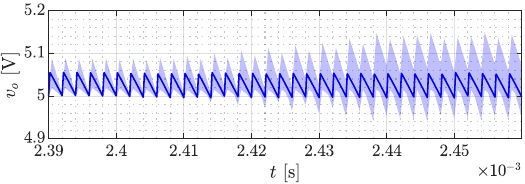}
    \end{minipage}
    \caption{EMPC robust performance: closed-loop system response $v_o(t)$ (nominal~\tikzref[color=blue, line width=0.75pt], uncertainty~\tikzreffill[fill=blue, fill opacity=0.25]) to a step line disturbance $v_\mathrm{in}(t)$.}
    \vspace{-0.5\baselineskip}
    \label{fig:buck_empc_sim_mc_line}
\end{figure*}

We study the response of the closed-loop system, i.e., the controlled output voltage $v_o(t)$, in presence of both load and line variations, i.e., injecting the disturbances $i_o(t)$ and $v_\mathrm{in}(t)$, respectively.
Fig.~\ref{fig:buck_empc_sim_mc_load} reports the closed-loop system response $v_o(t)$ in presence of a step load disturbance $i_o(t)$ applied at time $t_{i_o}$, with amplitude equal to $I_{o,\mathrm{max}}-\frac{\overline{V}_o}{R_L}$; such an amplitude is chosen so that the total output current jumps exactly to the worst case value $I_{o,\mathrm{max}}$. Fig.~\ref{fig:buck_empc_sim_mc_load}a reports the ceramic capacitor case with $t_{i_o}=\num{2.4e-03}\;\unit{s}$; Fig.~\ref{fig:buck_empc_sim_mc_load}b reports the electrolytic capacitor case with $t_{i_o}=\num{2e-03}\;\unit{s}$.
Fig.~\ref{fig:buck_empc_sim_mc_line} reports the closed-loop system response $v_o(t)$ in presence of a step line disturbance $v_\mathrm{in}(t)$ of amplitude to $10\;\unit{V}$ applied at time $t_{v_\mathrm{in}}$; such an amplitude is significantly high compared to the nominal line voltage $\overline{V}_\mathrm{in} = \num{50}\;\unit{V}$.
Fig.~\ref{fig:buck_empc_sim_mc_line}a reports the ceramic capacitor case with $t_{v_\mathrm{in}}=\num{2.8e-03}\;\unit{s}$; Fig.~\ref{fig:buck_empc_sim_mc_line}b reports the electrolytic capacitor case with $t_{v_\mathrm{in}}=\num{2.4e-03}\;\unit{s}$.

Overall, we observe that the EMPC policy~\eqref{eq:buck_empc_law_true}, controlling the Buck converter with parametric uncertainty, achieves consistently good performance across the entire range of admissible component values, demonstrating the robustness of the adopted methodology.
In general, we observe that parametric uncertainty leads to a slightly erroneous steady-state value, i.e., the reference output $\overline{V}_o = 5\;\unit{V}$ is not exactly tracked. Still, the error is always lower compared to the ripple voltage, which is acceptable for most applications.

Regarding disturbance rejection, very small settling times are achieved across all simulations. Such a settling time is comparable to just a few cycles of the PWM modulation frequency: on average, three cycles are sufficient to restore steady-state cyclostationary operation. 
In the case of load disturbance, the undershoot/overshoot strongly depends on the disturbance amplitude and the capacitor ESR value $R_{C_o}$, being significantly larger for the electrolytic capacitor case.
Specifically, we see that, when considering the electrolytic capacitor (i.e., higher $R_{C_o}$), the load disturbance undershoot becomes more pronounced, but the settling time improves.
As concerns line disturbance, we observe, as expected, a small steady-state tracking error due to the linearized model nonlinearity involving $V_\mathrm{in}$. However, this effect is always comparable to, or less than, the voltage ripple.

\subsection{Circuital Simulations} \label{sec:spice}

\begin{figure*}[t!]
    \centering
    \begin{minipage}[t]{0.32\linewidth}
        \centering
        \caption*{(a) Supply voltage $V_\mathrm{in} = 40\;\unit{V}$ ($v_\mathrm{in} = -10\;\unit{V}$)}
        \vspace{-0.5\baselineskip}
        \includegraphics[width=\linewidth]{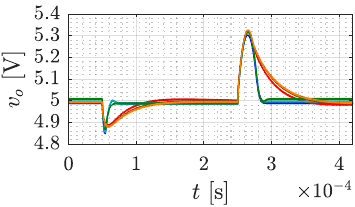}
    \end{minipage}
    \hfill 
    \begin{minipage}[t]{0.32\linewidth}
        \centering
        \caption*{(b) Supply voltage $V_\mathrm{in} = 50\;\unit{V}$ ($v_\mathrm{in} = 0\;\unit{V}$)}
        \vspace{-0.5\baselineskip}
        \includegraphics[width=\linewidth]{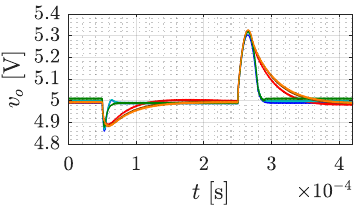}
    \end{minipage}
    \hfill 
    \begin{minipage}[t]{0.32\linewidth}
        \centering
        \caption*{(c) Supply voltage $V_\mathrm{in} = 60\;\unit{V}$ ($v_\mathrm{in} = 10\;\unit{V}$)}
        \vspace{-0.5\baselineskip}
        \includegraphics[width=\linewidth]{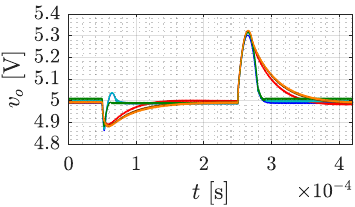}
    \end{minipage}
    
    \vspace{0.5\baselineskip} 
    \caption{EMPC and VMC response $v_o(t)$ to a \SI{10}{\ampere} pulse load disturbance $i_o(t)$, under different constant line disturbance $v_\mathrm{in}$ and resistive loads $R_L$: EMPC with  $R_L=1\;\unit{\ohm}$~\tikzref[color=spice-blue, line width=0.8pt], $R_L=3\;\unit{\ohm}$~\tikzref[color=spice-cyan, line width=0.8pt], $R_L=5\;\unit{\ohm}$~\tikzref[color=spice-green, line width=0.8pt]; VMC with $R_L=1\;\unit{\ohm}$~\tikzref[color=spice-red, line width=0.8pt], $R_L=3\;\unit{\ohm}$~\tikzref[color=spice-brown, line width=0.8pt], $R_L=5\;\unit{\ohm}$~\tikzref[color=spice-orange, line width=0.8pt]).}
    \label{fig:buck_empc_sim_spice}
\end{figure*}

\begin{figure*}[t!]
    \centering
    \begin{minipage}[t]{0.32\linewidth}
        \centering
        \caption*{(a) Supply voltage $V_\mathrm{in} = 40\;\unit{V}$ ($v_\mathrm{in} = -10\;\unit{V}$)}
        \vspace{-0.5\baselineskip}
        \includegraphics[width=\linewidth]{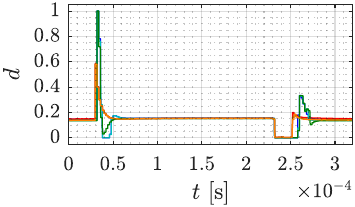}
    \end{minipage}
    \hfill 
    \begin{minipage}[t]{0.32\linewidth}
        \centering
        \caption*{(b) Supply voltage $V_\mathrm{in} = 50\;\unit{V}$ ($v_\mathrm{in} = 0\;\unit{V}$)}
        \vspace{-0.5\baselineskip}
        \includegraphics[width=\linewidth]{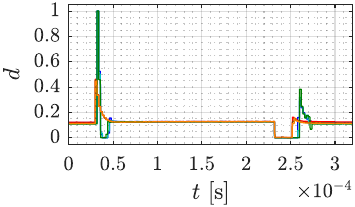}
    \end{minipage}
    \hfill 
    \begin{minipage}[t]{0.32\linewidth}
        \centering
        \caption*{(c) Supply voltage $V_\mathrm{in} = 60\;\unit{V}$ ($v_\mathrm{in} = 10\;\unit{V}$)}
        \vspace{-0.5\baselineskip}
        \includegraphics[width=\linewidth]{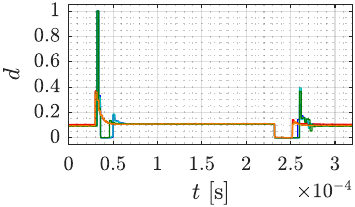}
    \end{minipage}
    
    \vspace{0.5\baselineskip} 
    \caption{EMPC and VMC duty cycle $d_k$ in response to a \SI{10}{\ampere} pulse load disturbance $i_o(t)$, under different constant line disturbance $v_\mathrm{in}$ and resistive loads $R_L$: EMPC with  $R_L=1\;\unit{\ohm}$~\tikzref[color=spice-blue, line width=0.8pt], $R_L=3\;\unit{\ohm}$~\tikzref[color=spice-cyan, line width=0.8pt], $R_L=5\;\unit{\ohm}$~\tikzref[color=spice-green, line width=0.8pt]; VMC with $R_L=1\;\unit{\ohm}$~\tikzref[color=spice-red, line width=0.8pt], $R_L=3\;\unit{\ohm}$~\tikzref[color=spice-brown, line width=0.8pt], $R_L=5\;\unit{\ohm}$~\tikzref[color=spice-orange, line width=0.8pt].}
    \label{fig:buck_empc_sim_spice_duty}
\end{figure*}


We assess the control performance of the analog EMPC circuit controlling the Buck converter through circuit-level simulations.
These simulations, carried out in LTSpice, are extremely accurate and allow us to demonstrate the practical feasibility of the proposed approach in real-world conditions. Through these simulations, we can also investigate the impact of circuit non-idealities on control performance.

In order to conduct the simulations, we select commercially-available components to implement the circuit (refer to Sec.~\ref{sec:empc_circ_implem} for more details).
For each component, the manufacturer model has been imported into the LTSpice scheme. This enables accurate transistor-level simulation results, which include the non-idealities of the selected components. Among these, we have the finite gain-bandwidth product, the offset currents and voltages, and the slew-rate limitations for OP-AMPs, and the response delay for comparators, MUX, and logic gates. 
In particular, the non-ideal component dynamics and propagation delays accurately reflect real-time implementation conditions. Under these realistic constraints, the analog controller achieves a total latency of approximately $1\,\unit{\micro\second}$, which is strictly lower than the design sampling period and switching period $T = 2\,\unit{\micro\second}$ ($f_\mathrm{sw} = 500\,\unit{kHz}$). This validates the circuit's real-time feasibility and demonstrates its capability for high-frequency operation. Furthermore, this latency represents a substantial improvement over the $6\,\unit{\micro\second}$ reported in~\cite{vichik_solving_2014} (enabled by the absence of slew-rate limitations; see Sec.~\ref{sec:comp_circuits}) as well as existing digital MPC implementations for Buck converters~\cite{cim25,mariethoz_comparison_2010,liu_fast_2018,albira_adaptive_2021}. Finally, while implementing this controller using the approach in~\cite{vichik_solving_2014} would require $8$ OP-AMPs, the proposed architecture achieves it with only $2$ (see Sec.~\ref{sec:comp_circuits}).

Alternative MPC implementation approaches, such as~\cite{WillsTCST12_fpga,vichik_solving_2014} cannot operate at the target $\SI{500}{\kilo\hertz}$ switching frequency. Simulations under the conditions in Sec.~\ref{sec:simulink} at $\SI{100}{\kilo\hertz}$, a frequency compatible with the $\SI{6}{\micro\second}$ propagation delay of~\cite{vichik_solving_2014}, show that although transient performance is comparable to the $\SI{500}{\kilo\hertz}$ case, the steady-state output voltage ripple quintuples, in accordance with the theory in~\cite{book-erickson-2020}. As this elevated ripple is unacceptable for the target application, these lower-frequency waveforms are omitted. Instead, the quantitative comparison focuses on voltage-mode control (VMC), a well-established strategy compatible with high-frequency operation.

To test different operating conditions, we conduct $9$ simulations, considering $3$ values for the resistive load $R_L \in \{1\;\unit{\ohm}, \; 3\;\unit{\ohm}, \; 5\;\unit{\ohm}\}$ and $3$ values for the supply voltage $V_\mathrm{in} \in \{40\;\unit{V}, \; 50\;\unit{V}, \; 60\;\unit{V}\}$. The latter model a constant line disturbance $v_\mathrm{in} \in \{-10\;\unit{V}, \; 0\;\unit{V}, \; 10\;\unit{V}\}$. To evaluate load disturbance rejection, we consider an output current pulse $i_o(t)$ with an amplitude $10\;\unit{A}$ and duration of $0.2\;\unit{ms}$.
Fig.~\ref{fig:buck_empc_sim_spice} reports the closed-loop system response, i.e., the controlled output voltage $v_o(t)$. %
For each simulation, the response is compared with that obtained using standard VMC~\cite{book-erickson-2020} implemented in LTSpice.
Detailed VMC controller design, implementation, and component specifications can be found in~\cite{gab25_estimator}, which utilizes the same case study.
Results indicate that, with the selected components, the impact of circuit non-idealities is negligible, as the circuit-level system responses are mostly overlapping with those obtained through the high-level simulations in Sec.~\ref{sec:simulink}.
In terms of load disturbance rejection, the analog EMPC circuit significantly outperforms standard VMC. At $t=\num{0.5e-04}\;\unit{s}$ ($i_o$ rising edge), the undershoot is comparable between EMPC and VMC, being $2.4\%$ for VMC and $2.6\%$ for EMPC; instead, the settling time is $2.5\;\unit{\micro\second}$ for EMPC, significantly lower than the $10\;\unit{\micro\second}$ of VMC. Similarly, during the recovery phase at $t=\num{2.5e-04}\;\unit{s}$ ($i_o$ falling edge), we observe identical average overshoots of $6.2\%$ and the settling time is $42\;\unit{\micro\second}$ for EMPC, while VMC exceeds $200\;\unit{\micro\second}$. These results are consistent across the various considered supply voltages and load configurations.

Fig.~\ref{fig:buck_empc_sim_spice_duty} shows the duty cycle response to the pulse load disturbance. As expected, EMPC saturates the duty command to rapidly reject the disturbance, whilst the VMC response is more conservative across all considered operating conditions.

In all simulations, both EMPC and VMC exhibit a small steady-state tracking error, which, in either case, is at most $10\,\unit{mV}$.
For EMPC, this error arises from the joint effects of the uncertain load value $R_L$, the model linearization error, and the approximation introduced by rounding the resistors used to implement the MPC to standard E-series values.
Conversely, VMC benefits from an inherent integral action that theoretically ensures zero steady-state error, but the finite gain of the OP-AMP used in its implementation leads to a small non-zero steady-state error.

\section{Conclusions} \label{sec:conclusion}

This paper introduced a general methodology for implementing explicit model predictive control policies as fully analog electronic circuits.
The proposed approach leverages the piecewise-affine structure of the control policy to map the control law into an analog architecture and employs tailored complexity-reduction strategies to minimize the number of components. This enables real-time operation within fast sampling rates and avoids the overhead associated with digital implementations. 

Our approach was applied to the control of DC-DC Buck converters, achieving effective rejection of load and line disturbances. We theoretically analyzed the robustness of closed-loop stability to bounded disturbances, leveraging contraction theory. The EMPC analog circuit was validated through extensive Monte Carlo and circuit-level simulations, demonstrating excellent disturbance rejection performance across a wide range of uncertainties.

Future work will explore the application of the approach to other plants that require fast sampling, such as different DC-DC converter topologies.

\bibliographystyle{ieeetr}
\bibliography{references}
 
\appendix

\section{Reformulation of the MPC problem in standard QP form}
\label{app:qp_formul}
This section shows how the MPC problem~\eqref{eq:mpc_op} can be rewritten in a standard and more compact QP form, comprising only $\hat{u}$ as decision variables. The adopted procedure follows along the same lines as in~\cite{bemporad2000explicit}.

We eliminate the constraints~\eqref{eq:mpc_op_b},~\eqref{eq:mpc_op_c} and rewrite Eq.~\eqref{eq:mpc_op_a} as a function of $\hat{u}$ and $p_k$ only. First, to eliminate the variables $\hat{x}$ and $\hat{y}$ from problem~\eqref{eq:mpc_op}, by recursively applying Eq.~\eqref{eq:mpc_op_b}, we have
\begin{align}
    \hat{x}_i = A^i x_k + \sum_{j=0}^{i-1} A^{i-j-1}(B \hat{u}_i + B_\nu \nu_k + b).
\end{align}
Then, $\hat{x}$ is expressed as a function of $\hat{u}$ and $p_k$ by
\begin{align} \label{eq:mpc_to_qp_calc_1}
    \hat{x} = \begin{bmatrix} \Phi & \Gamma_\nu \end{bmatrix} p_k + \Gamma \hat{u} + \gamma,
\end{align}
where
\begin{subequations} \label{eq:mpc_to_qp_calc_2}
\begin{align}
    \Phi &= \begin{bmatrix}
        I_n \\ \sum_{i=1}^{N_p} e_{N_p}^{(i)} \otimes A^i
    \end{bmatrix}, \\
    \Gamma &= \begin{bmatrix}
        \bm{0}_{n \times N_p n_u} \\ I_{N_p} \otimes B + \sum_{i=1}^{N_p-1} E_{N_p}^{(i)} \otimes A^i B
    \end{bmatrix}, \\
    \Gamma_\nu &= \begin{bmatrix}
        \bm{0}_{n \times n} \\ \sum_{i=1}^{N_p} \left( e_{N_p}^{(i)} \otimes \sum_{j=0}^{i-1} A^j \right) 
    \end{bmatrix} B_\nu, \\
    \gamma &= \begin{bmatrix}
        \bm{0}_{n \times n} \\ \sum_{i=1}^{N_p} \left( e_{N_p}^{(i)} \otimes \sum_{j=0}^{i-1} A^j \right)
    \end{bmatrix} b.
\end{align}
\end{subequations}

By Eq.~\eqref{eq:mpc_op_c}, we can also rewrite
\begin{align} \label{eq:mpc_to_qp_calc_3}
    \hat{y} = \overline{C}\hat{x} + \overline{D}\hat{u} + \overline{D}_\nu \nu_k + \overline{d},
\end{align}
where
\begin{subequations} \label{eq:mpc_to_qp_calc_4}
    \begin{alignat}{3}
        \overline{C} &= \begin{bmatrix} I_{N_p} \otimes C & \bm{0}_{n_y \times n} \end{bmatrix}, &\quad \overline{D} &= I_{N_p} \otimes D, \\
        \overline{D}_\nu &= I_{N_p} \otimes D_\nu, &\quad \overline{d} &= \bm{1}_{N_p} \otimes d.
    \end{alignat}
\end{subequations}

Next, by defining 
\begin{align}
    & \overline{Q} = I_{N_p} \otimes Q, \nonumber \\
    & \overline{R} = I_{N_p} \otimes R, \quad \overline{R}_\Delta = M \otimes R_\Delta, \nonumber \\
    & M = \diag([1, 2 \cdot \bm{1}_{N_p-2}^\top, 1]^\top) - E_{N_p}^{(1)} - E_{N_p}^{(1)\,\top}, \nonumber \\
    & \overline{y}_{r} = \bm{1}_{N_p} \otimes y_r,
\end{align}
we rewrite the cost function $J(\hat{y},\hat{u})$ in Eq.~\eqref{eq:mpc_op_a} in the following compact form:
\begin{align} \label{eq:mpc_op_dost_compact}
    J(\hat{y},\hat{u}) = \|\hat{y} - \overline{y}_r \|_{\overline{Q}}^2 + \| \hat{u} \|_{\overline{R}+\overline{R}_\Delta}^2.
\end{align}

Replacing Eqs.~\eqref{eq:mpc_to_qp_calc_1} and~\eqref{eq:mpc_to_qp_calc_3} into Eq.~\eqref{eq:mpc_op_dost_compact} yields 
\begin{align} \label{eq:mpc_qp_cost_calc}
    & J(\hat{y},\hat{u}) = J_u(\hat{u}) = \frac{1}{2}\hat{u}^\top H \hat{u} + (F p_k + c)^\top \hat{u},
\end{align}
where
\begin{subequations}
    \begin{align}
        H &= (\overline{C}\Gamma + \overline{D})^\top \overline{Q} (\overline{C}\Gamma + \overline{D}) + \overline{R} + \overline{R}_\Delta, \\
        F &= (\overline{C}\Gamma + \overline{D})^\top \overline{Q} \, \overline{C} \begin{bmatrix}
            \Phi & \overline{C}\Gamma_\nu + \overline{D}_\nu
        \end{bmatrix}, \\
        c &= (\overline{C}\Gamma + \overline{D})^\top \overline{Q} (\overline{C} \gamma + \overline{d} - \overline{y}_r).
    \end{align}
\end{subequations}

Finally, we also express the linear constraints~\eqref{eq:mpc_op_d} as a function of $\hat{u}$ and $p_k$ only.
First, we compact Eq.~\eqref{eq:mpc_op_d} as
\begin{align} \label{eq:mpc_to_qp_calc_5}
    \overline{H}_x \hat{x} \leq h_x, \quad \overline{H}_u \hat{u} \leq h_u
\end{align}
where
\begin{subequations} \label{eq:mpc_to_qp_calc_6}
    \begin{alignat}{3}
        \overline{H}_x &= \begin{bmatrix} I_{N_p} \otimes H_x & \bm{0} \end{bmatrix}, &\quad \overline{h}_x &= \bm{1}_{N_p} \otimes h_x, \\
        \overline{H}_u &= I_{N_p} \otimes H_u, &\quad \overline{h}_u &= \bm{1}_{N_p} \otimes h_u.
    \end{alignat}
\end{subequations}
Then, replacing Eq.~\eqref{eq:mpc_to_qp_calc_1} into Eq.~\eqref{eq:mpc_to_qp_calc_5} yields the constraints
\begin{align}
    G \hat{u} \leq w + Kp_k,
\end{align}
where
\begin{align}
    G &= \begin{bmatrix}
        \overline{H}_x \Gamma \\
        \overline{H}_u
    \end{bmatrix}, \quad w = \begin{bmatrix}
        \overline{h}_x - \overline{H}_x \gamma \\
        \overline{h}_u
    \end{bmatrix}, \nonumber \\
    K &= \begin{bmatrix}
        -\overline{H}_x \Phi & -\overline{H}_x \Gamma_\nu \\
        \bm{0} & \bm{0}
    \end{bmatrix}.
\end{align}

Overall, the original MPC problem~\eqref{eq:mpc_op} is equivalent to the following QP:
\begin{subequations} \label{eq:mpc_qp_app}
    \begin{align}
        \min_{\hat{u}} \quad & \frac{1}{2}\hat{u}^\top H \hat{u} + (F p_k + c)^\top \hat{u} \label{eq:mpc_qp_a_app} \\
        \textrm{s.t.} \quad & G \hat{u} \leq w + Kp_k. \label{eq:mpc_qp_b_app}
    \end{align}
\end{subequations}

\end{document}